\documentclass{aa}  
\usepackage{natbib}

\usepackage{color}

\usepackage{graphicx}
\usepackage{txfonts}
\usepackage[switch]{lineno} % 使用switch选项
\begin{document}
\title{Evolution of low-mass He stars and implications for electron-capture supernova formation in close binaries}

\titlerunning{Evolution of low-mass He stars and implications for the formation of ECSN in close binaries}

\author{Jun-Qian Li\inst{1,2}
        \and
        Ying Qin\inst{1,2} 
         \and
        Zi-Yuan Wang\inst{1,2} 
         \and           
         Qing-Wen Tang\inst{3}
          \and
        Han-Feng Song\inst{4}
           \and
        Georges Meynet\inst{5,6}   
        }
\authorrunning{Li et al.}
\institute{Department of Physics, Anhui Normal University, Wuhu, Anhui, 241002, China \\
        \email{yingqin@ahnu.edu.cn}         
        \and 
        Center for Astrophysics and Astronomical Technology, Anhui Normal University, Wuhu, Anhui 241002, China
        \and
        Department of Physics, School of Physics and Materials Science, Nanchang University, Nanchang 330031, China
        \and     
        College of Physics, Guizhou University, Guiyang city, Guizhou Province, 550025, China
        \and 
        Département d’Astronomie, Université de Genève, Chemin Pegasi 51, 1290 Versoix, Switzerland
        \and 
        Gravitational Wave Science Center (GWSC), Université de Genève, 24 quai E. Ansermet, 1211 Geneva, Switzerland
}
% \abstract{}{}{}{}{} 
% 5 {} token are mandatory
 \abstract
   {The evolution of low-mass helium (He) stars ($\sim 2.5-5\,M_\odot$) with neutron-star (NS) companions in close binaries has been extensively studied using detailed stellar and binary evolution models; however, the combined effects of rotation and tidal interaction on their evolution and final fate have not yet been systematically explored.}
   {We investigate how rotation, mass transfer, and tidal interactions influence the evolution of low-mass helium stars and the conditions leading to electron-capture supernovae (ECSNe) formation. We also predict the properties of the resulting neutron stars, including their spin periods, rotational energies, and internal magnetic field strengths.}
   {We perform detailed stellar structure and binary evolution calculations that account for mass loss, internal differential rotation, and tidal interactions between He stars and NS companions, and systematically map the initial binary parameter space leading to the formation of ECSNe.}
   {We first find that rotation has only a modest impact on the overall evolution of low-mass He stars. With a parameter-space study of low-mass He stars in close binaries, we find that ECSNe occur within a narrow initial He-star mass range of $2.42 - 2.67\, M_\odot$ at solar metallicity ($Z_\odot$) and $2.37 - 2.62\, M_\odot$ at $0.01\, Z_\odot$. The resulting NSs via ECSNe have spin periods of $7.7 - 83.8\, ms$ and magnetic fields of order $10^{12}$ G. Their rotational energies span in a range of $2.6\times10^{48}$--$2.5\times10^{50}$ erg ($2.7\times10^{48}$--$2.6\times10^{50}$ erg for $0.01\, Z_\odot$), although these values can be highly reduced by efficient angular-momentum transport mechanisms such as the Spruit--Tayler dynamo.}
   {Detailed binary evolution calculations have demonstrated that the evolutionary fate of low-mass He stars in close binaries is highly sensitive to the initial orbital period. Binaries with shorter orbital periods undergo Roche-lobe overflow at earlier stages of evolution, resulting in stronger binary interactions. Comparison with Galactic double NS systems shows that most observed systems can be reproduced in the eccentricity--orbital-period plane when relatively large natal kicks are assumed.}


\maketitle

\section{Introduction}\label{sect1}
Low-mass helium (He) stars in close binaries with neutron-star (NS) companions are widely regarded as the immediate progenitors of double neutron star (DNS) systems in the standard isolated binary evolution scenario \citep[e.g.,][]{Tauris2017}. The subsequent evolution of the He star determines the nature of the second supernova, the properties of the newly formed NS, and ultimately the orbital characteristics of the resulting DNS system. Consequently, understanding the evolution and final fate of low-mass He stars is essential for constraining the formation and properties of DNS systems.

Detailed binary evolution calculations of low-mass He stars with NS companions in different parameter spaces have been carried out in a number of previous studies. \cite{Dewi2002} were the first to investigate the evolution of He stars with masses of $1.5 - 6.7\, M_\odot$ orbiting a 1.4 $M_\odot$ NS, focusing on Case BA (mass transfer initiates during core helium burning phase) and Case BB (mass transfer initiates during shell helium burning phase) mass transfer as well as the final fate of the He star as a function of its mass and orbital period. They subsequently examined the late evolutionary stages of He stars with masses of $2.8 - 6.4\, M_\odot$ in wider orbits with a 1.4 $M_\odot$ NS companion, exploring the possible outcomes of these systems \citep{Dewi2003}. Later, \cite{Ivanova2003} calculated the evolution of binaries containing He stars with masses of $2.5 - 6\, M_\odot$ and a 1.4 $M_\odot$ NS companion to investigate the formation of double NS systems. Focusing on lower-mass He stars ($2.5 - 3.5\, M_\odot$) with NS companions, \cite{Tauris2015} presented a systematic study of the evolutionary pathways leading to ultra-stripped SNe. It was pointed out by \cite{Wu2022} that lower-mass ($2 - 3\, M_\odot$) He stars could undergo substantial envelope expansion during the later core neon and oxygen burning, leading to extremely high mass transfer rates ($ > 10^{-2}$ $M_\odot$ $\rm yr^{-1}$). More recently, \cite{Guo2024} investigated the formation of electron-capture supernovae (ECSNe) in He star-NS binaries and their role in the formation of double NS systems. Notably, they also explored the impact of a residual hydrogen envelope on helium star (He-star) evolution. He stars retaining a hydrogen envelope experience more pronounced radial expansion, which promotes stronger mass transfer to the NS companion and leads to more efficient recycling of the NS.

Among these outcomes, ECSNe are of particular interest because they are expected to produce relatively low explosion energies and weak natal kicks, making them a promising formation channel for DNS systems \citep[e.g.,][]{Podsiadlowski2004,Tauris2015}. Electron-capture collapse occurs when a degenerate oxygen–neon (ONe) core approaches the Chandrasekhar mass and becomes unstable owing to electron captures on nuclei such as $^{24}$Mg and $^{20}$Ne \citep{Miyaji1980,Nomoto1984,Nomoto1987}. However, the evolutionary pathways leading to ECSNe in close binaries remain uncertain, owing to the complex interplay between binary interaction and internal stellar evolution.

In addition to mass transfer, the evolution of low-mass He stars is influenced by rotation and tidal interaction. Tidal torques exchange angular momentum between the stellar spin and the orbit, affecting both the rotational evolution of the He star and the orbital evolution of the binary. Rotation can further induce the transport of angular momentum and chemical species within the stellar interior, thereby modifying the growth of the inner core. Since ECSN progenitors occupy a relatively narrow region of parameter space \cite[e.g.,][]{Tauris2015,Guo2024}, these processes may influence the conditions for electron-capture collapse and the properties of the resulting neutron stars. Nevertheless, the combined effects of mass transfer, rotation, and tidal interaction on the evolution of low-mass He stars and the formation of ECSN progenitors have not yet been systematically explored.

In this work, we perform detailed binary evolution calculations of low-mass He stars with NS companions in close binaries to investigate how mass transfer, rotation, and tidal interaction influence their inner structure and the conditions leading to ECSN formation. The structure of this paper is as follows. In Section~\ref{sect2}, we describe the main methods adopted in the stellar and binary evolution models. In Section~\ref{sect3}, we present the impacts of rotation on the evolution of single low-mass He stars, and their detailed evolution in close binaries (including properties of two components, parameter space analysis, and connection to the observed DNSs in the Milky Way). Finally, we present our conclusions with discussion in Section~\ref{sect4}.

\section{Methods}\label{sect2}
We performed stellar evolution and binary modeling using the release version \texttt{mesa-r15140} of the Modules for Experiments in Stellar Astrophysics (\texttt{MESA}) stellar evolution code \citep{Paxton2011,Paxton2013,Paxton2015,Paxton2018,Paxton2019, Jermyn2023}. Our low-mass He star zero-age main sequence models were constructed following the methodology outlined in earlier studies \citep[e.g.,][]{Fragos2023,lv2023}. In this work, we adopted a solar metallicity of $Z_{\odot} = $ 0.0142 \citep{Asplund2009}.

We modeled convection using the mixing-length theory \citep{MLT1958}, adopting a mixing-length parameter of $\alpha_{\rm mlt}=1.93$, following the prescription adopted in the POSYDON framework \citep{Fragos2023}. Convective boundaries were determined according to the Ledoux criterion, and overshooting was implemented using a step prescription with $\alpha_{\rm p} = 0.1\, H_{\rm p}$, where $H_{\rm p}$ is the pressure scale height at the Ledoux boundary. Semiconvection was included in the He star models following \cite{Langer1983}, with an efficiency parameter of $\alpha_{\rm sc}=1.0$. Nucleosynthesis was calculated using the \texttt{approx21.net} reaction network.

We treated rotational mixing and angular momentum transport as diffusive processes \citep{Heger2000}, incorporating the effects of the Goldreich–Schubert–Fricke instability, Eddington–Sweet circulations, as well as secular and dynamical shear mixing. The efficiency of diffusive element mixing was set to $f_{\rm c} = 1/30$, following \cite{Chaboyer1992,Heger2000}. To account for the sensitivity of the $\mu$-gradient to rotationally induced mixing, we mitigated its impact by multiplying $f_\mu = 0.05$, as recommended by \cite{Heger2000}. Additionally, the rotationally enhanced mass loss based on prescription as in \cite{Heger1998} and \cite{Langer1998} was included. Notably, we did not include gravity-darkening effects, as discussed in \cite{Maeder2000}.

We applied the dynamical tides to He stars with radiative envelopes, following the framework of \cite{Zahn1977}. The synchronization timescale was computed using the prescriptions of \cite{Zahn1977}, \cite{Hut1981}, and \cite{Hurley2002}, while the tidal torque coefficient $E_2$ was adopted from the updated fitting formula of \cite{Qin2018}. Owing to previously identified inconsistencies in the implementation of the synchronization timescale \citep{Sciarini2024}, we adopted the implementation described by \cite{Qin2024_gap}. Stellar wind mass loss was treated under the Jeans mode assumption, in which the wind carries away the specific angular momentum of the mass-losing star. Unless otherwise stated, all binary evolution calculations are terminated at central carbon depletion of the He star. For mass transfer from He stars onto NS, we assume Eddington-limited accretion. For simplicity, we adopt mdot$_{-}$scheme = ``Roche$_{-}$lobe'' such that mass transfer is switched off and the system is considered to be detached when the donor remains inside its Roche lobe.

In \texttt{MESA}, the standard mass-loss prescriptions are unified into the Dutch wind scheme, which adopts the rates of \citet{Vink2001} for hot, hydrogen-rich stars, \citet{deJager} for cool stars, and \citet{Nugis2000} for hot stars that have lost their hydrogen envelopes. Thus, we adopt the mass-loss wind prescription for He stars in this work following \citet{Nugis2000} (hereafter NL2000) as follows:
\begin{equation}\label{}
\log \left(\frac{\dot{M}_{\rm NL2000}}{M_\odot\,\mathrm{yr}^{-1}}\right) = -11.00 + 1.29 \log(L/L_\odot) + 1.7 \log Y + 0.5 \log Z_{\rm cur},
\end{equation}

\noindent
where $Z_{\rm cur}$ is the current metallicity and $Y$ is the He abundance. As suggested in \cite{Hu2022}, a scaling factor of $\eta =$ 2/3 is adopted to match the recently updated modeling of He star winds \citep{Higgins2021}.

In the present work, we also adopt the alternative wind prescription proposed by \citet{Vink2017} (hereafter V2017), which is more suitable for low-mass He stars with masses of $2-60\, M_\odot$:

\begin{equation}
\log \left(\frac{\dot{M}_{\rm V2017}}{M_\odot\,\mathrm{yr}^{-1}}\right) = -13.3 + 1.36 \log(L/L_\odot) + 0.61 \log(Z_{\rm cur}/Z_\odot).
\end{equation}

Low-mass He stars within a certain mass range are expected to end their lives as ECSNe. In this work, following \citet{Tauris2015,Fragos2023}, we define ECSN progenitors as models with carbon–oxygen (CO) core masses in the range of $1.37-1.43\, M_\odot$ at the end of central carbon depletion.

\section{Results}\label{sect3}
\subsection{Evolution of Single low-mass He stars}
In this section, we first present detailed stellar evolution calculations of low-mass He stars ($2.5-5\,M_\odot$) using the V2017 wind prescription at two inital metallicities ($Z_\odot$ and $0.01\,Z_\odot$ ). We then investigate the impact of stellar rotation on their evolution.

\begin{figure}[h]
     \centering
     \includegraphics[width=0.45\textwidth]{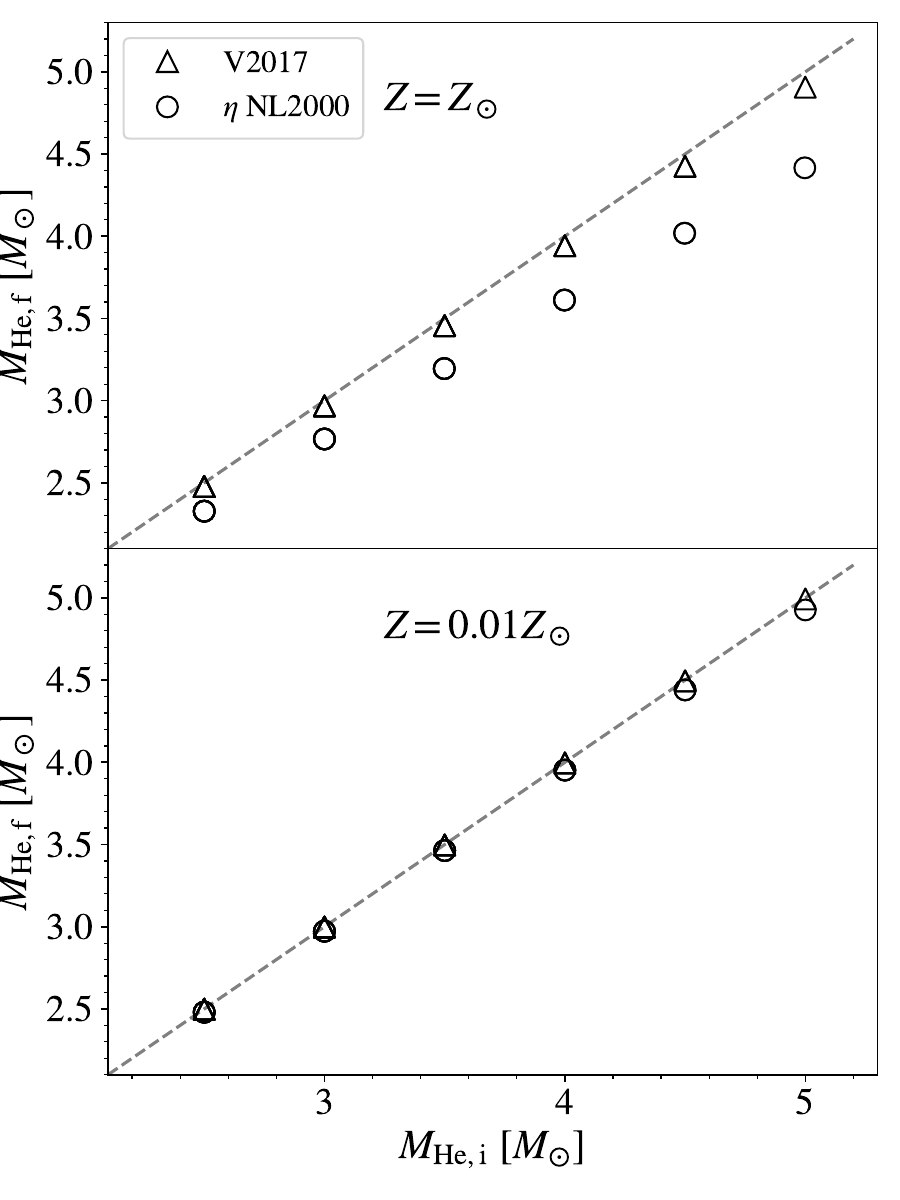}
     \caption{Final masses of single He stars without rotation at carbon depletion versus their initial masses. Different symbols represent different wind prescriptions, namely triangles for V2017 and circles for NL2000 (multiplied by a scaling factor of $\eta$ = 2/3). Two metallicities are considered: the upper panel shows $Z = Z_\odot$, while the lower panel corresponds to $Z = 0.01\,Z_\odot$. The dashed line marks $M_{\rm He,f} = M_{\rm He,i}$.}
     \label{M_wind} 
\end{figure}

\begin{figure}[h]
     \centering
     \includegraphics[width=0.49\textwidth]{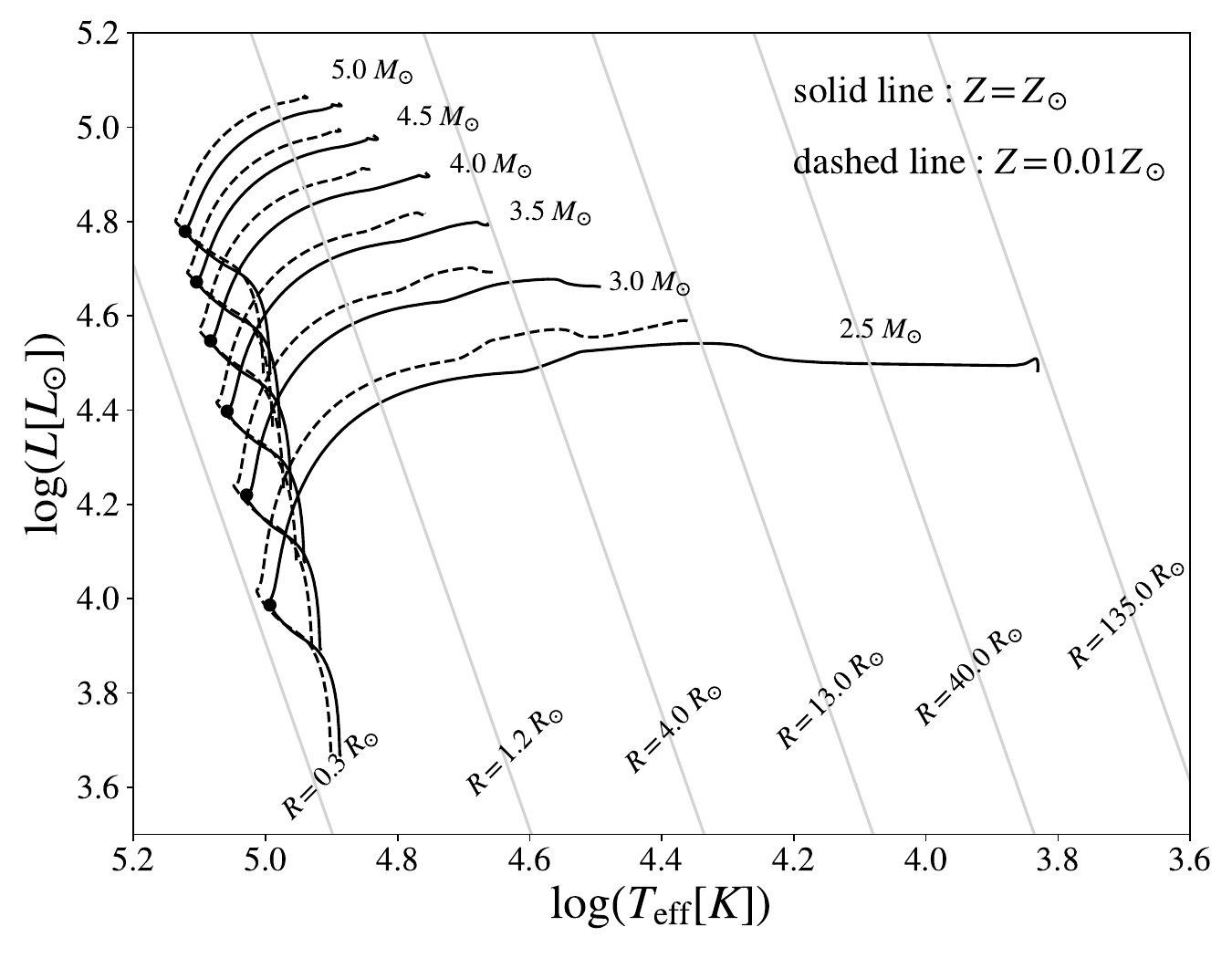}
     \caption{HR diagrams of single He-star models (without rotation) with initial masses ranging from 2.5 to 5 $M_\odot$. Solid lines represent models with solar metallicity ($Z=Z_\odot$), while dashed lines denote low-metallicity models of $Z=0.01\,Z_\odot$. The black dots indicate the stage of central helium depletion in the non-rotating models.}
     \label{HR} 
\end{figure}

In Figure~\ref{M_wind}, we present final masses of He stars at carbon depletion versus their initial masses (namely zero-age helium main sequence) using different wind prescriptions and metallicities. In the upper panel (solar metallicity), we note that the wind mass loss for V2017 is weaker when compared with NL2000, especially for higher initial masses. However, at a lower metallicity (0.01 $Z_\odot$), He stars are found to have nearly no mass loss for both prescriptions. With the wind prescription of V2017, we further show the Hertzsprung-Russell (HR) diagrams in Figure~\ref{HR} for He-star models at two different metallicties from the onset of the core helium burning to carbon depletion. First, He stars with lower initial metallicity are more luminous. This is because lower metallicity reduces opacity, producing more compact and hotter stellar structures. Second, He stars with initially lower masses expand more significantly when evolving off the main sequence. Therefore, lower-mass He stars are expected to have more easily mass exchange with their companions, especially in close binary systems.

As previously shown by \cite{Zhang2023}, rotating He stars undergo enhanced wind-driven mass loss due to rotational effects \citep{Langer1998}. We further investigate how the initial rotation velocity influences the evolution of low-mass He stars. The critical rotational frequency is given as follows:

\begin{equation}
\omega_{\rm crit}^2 = (1-\frac{L}{L_{\rm Edd}}) \frac{GM}{R^3},
\end{equation}
where $G$ is the gravitational constant, $M$ and $R$ denote the total mass and radius of the star, respectively, and $L$ is the stellar luminosity. $L_{\rm Edd}$ represents the Eddington luminosity, defined as follows:

\begin{equation}
L_{\rm Edd}=\frac{4\pi G M c}{\kappa},
\end{equation}
where $c$ is the speed of light in vacuum and $\kappa$ is the opacity that contributes from pure electron scattering, i.e., $\kappa = 0.2(1+X) cm^2g^{-1}$. In a pure He star, the hydrogen mass fraction $X = 0$. 

\begin{figure}[h]
     \centering
     \includegraphics[width=0.45\textwidth]{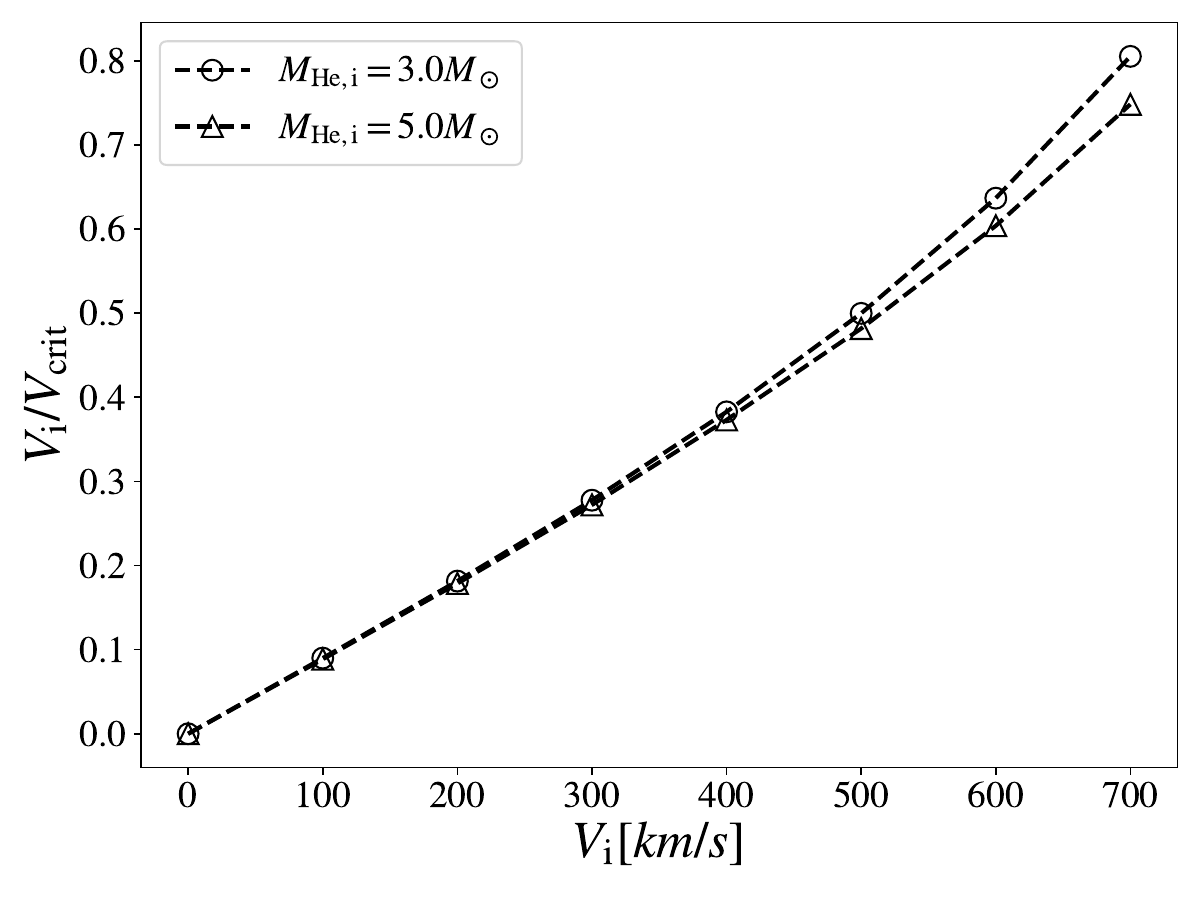}
     \caption{Ratio of the surface velocity to the critical velocity, $V/V_{\rm crit}$, as a function of the initial velocity for $5\,M_\odot$ (triangles) and $3\,M_\odot$ (circles) He stars at solar metallicity. The models are presented here for He stars at the evolutionary stage of central helium depletion.}
     \label{v_crit} 
\end{figure}

For the 3 and $5\,M_\odot$ He-star models (Unless otherwise specified, solar metallicity is assumed for all the models), we adopt initial rotational velocities ranging from $0$ to $700\,km/s$. Figure~\ref{v_crit} shows the ratio of the surface velocity to the critical velocity ($V_i/V_{\rm crit}$) as a function of the initial velocity for both models at the evolutionary stage of central helium depletion. Notably, $V_i/V_{\rm crit}$ is slightly higher for the lower initial mass, especially at initial velocities exceeding $400\,km/s$. 
This is because lower-mass He stars typically have a lower value of $M/R$ and thus lower critical velocities ($V_{\rm crit} \propto \sqrt{GM/R}$).

\begin{figure}[h]
     \centering
     \includegraphics[width=0.49\textwidth]{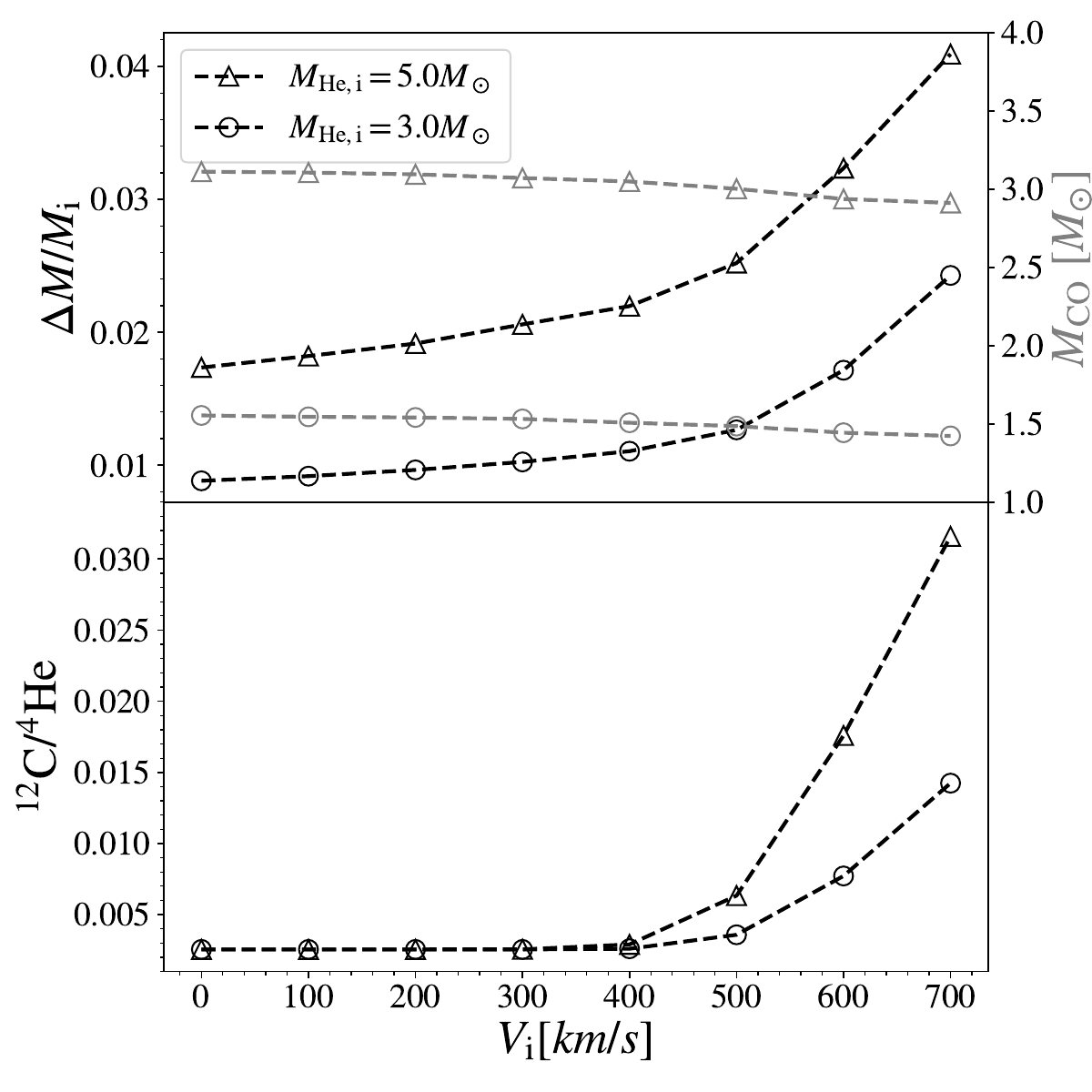}
     \caption{Relative mass change (left y axis of the upper panel), carbon-oxygen core mass (right y axis of the upper panel), and surface carbon-to-helium ratio (bottom panel) at the time of central helium depletion versus initial rotation. Dashed lines with triangles denote the $5\,M_\odot$ models and those with circles refer to the $3\,M_\odot$ models.}
     \label{M3_M5} 
\end{figure}

Figure~\ref{M3_M5} illustrates the relative mass loss, carbon-oxygen core mass, and surface carbon-to-helium ratio at central helium depletion. For a He star with $M_i = 3\,M_\odot$, the relative mass loss increases only slightly with increasing initial rotation, from $\sim\, 0.009$ ($V_{\rm i} = 0$ $km/s$) to $\sim\, 0.024$ ($V_{\rm i} = 700$ $km/s$). This trend becomes more pronounced for more massive He stars (see triangles). In addition, the carbon-oxygen core mass decreases slightly with increasing initial rotation, which can be attributed to the reduction in the total He-star mass at higher rotation rates. Meanwhile, the surface carbon-to-helium ratio remains nearly unchanged for $V_{\rm i} \leqslant 400$ $km/s$, and increases at higher rotation rates (see the bottom panel). This is because rotationally induced chemical mixing becomes more efficient in more massive He stars. Overall, we find that rotation exerts a slight effect on the extra mass-loss and chemical mixing of low-mass He stars compared to more massive He stars \cite[e.g.,][]{Qin2023}.

\subsection{Evolution of low-mass He stars in close binary systems}
Low-mass He stars in close binaries are expected to undergo mass transfer onto their companions \citep[e.g.,][]{Dewi2002,Dewi2003,Ivanova2003,Tauris2012,Tauris2015,Wang2024,Guo2024,Qin2024}, owing to their substantial radial expansion after core He exhaustion. We perform detailed binary evolution calculations for NS-He star systems with different initial orbital periods ($P_{\rm i}$ = 0.06, 0.35, and 2.18 d), aiming to investigate how the initial orbital period affects the mass-transfer phases, as well as the properties of the recycled NS and the second-born NS.

\subsubsection{Mass transfer and its impact on the recycled neutron star}
\begin{figure}[h]
     \centering
     \includegraphics[width=0.49\textwidth]{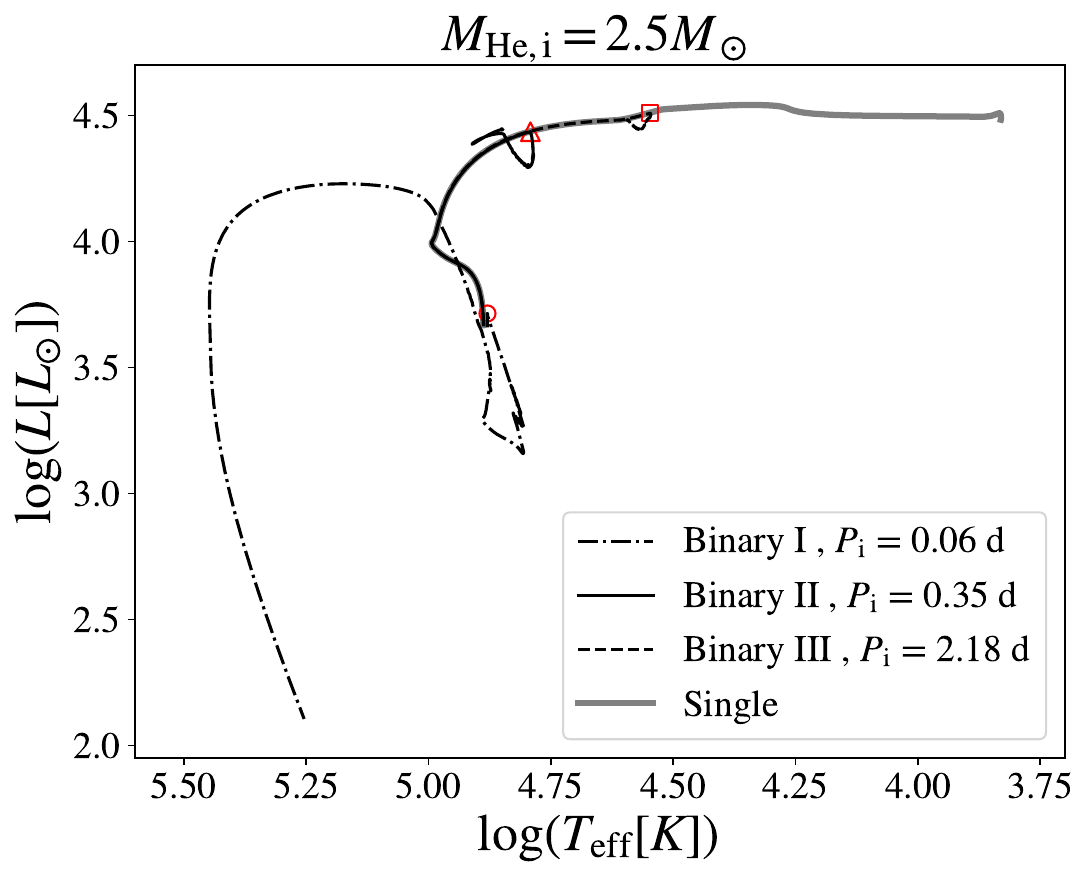}
     \caption{HR diagrams of $2.5\,M_{\odot}$ He-star models at solar metallicity. The black dash-dotted, solid, and dashed lines represent binary evolution sequences with initial orbital periods of $P_{\rm i} = 0.06$, 0.35, and 2.18 d, respectively, while the solid gray line denotes the corresponding single He-star model without rotation. Symbols mark the onset of different mass-transfer cases: red circles for Case BA, red triangles for Case BB, and red squares for Case BC (mass transfer initiates after the shell helium burning phase).}
     \label{HR_b} 
\end{figure}

Figure~\ref{HR_b} compares the HR diagram of a single $2.5\,M_\odot$ He star with that of He stars in close binaries spanning a range of initial orbital periods. As shown in Figure~\ref{HR_b}, the single He star remains compact ($R \sim 0.3\, R_\odot$) during the core helium burning phase, but undergoes substantial expansion after core He exhaustion, reaching $R \sim 135.0\,R_\odot$ by the stage of central carbon depletion. We further investigate the evolution of a $2.5\,M_\odot$ He star in a close binary system with a $1.35\,M_\odot$ NS companion for different initial orbital periods. The evolutionary tracks differ markedly from the single-star case, with both the onset and extent of the deviation depending on the initial orbital period. In particular, systems with shorter initial periods depart earlier from the single He-star track. These deviations are primarily driven by mass loss associated with the onset of Roche-lobe overflow. 

\begin{figure}[h]
     \centering
     \includegraphics[width=0.49\textwidth]{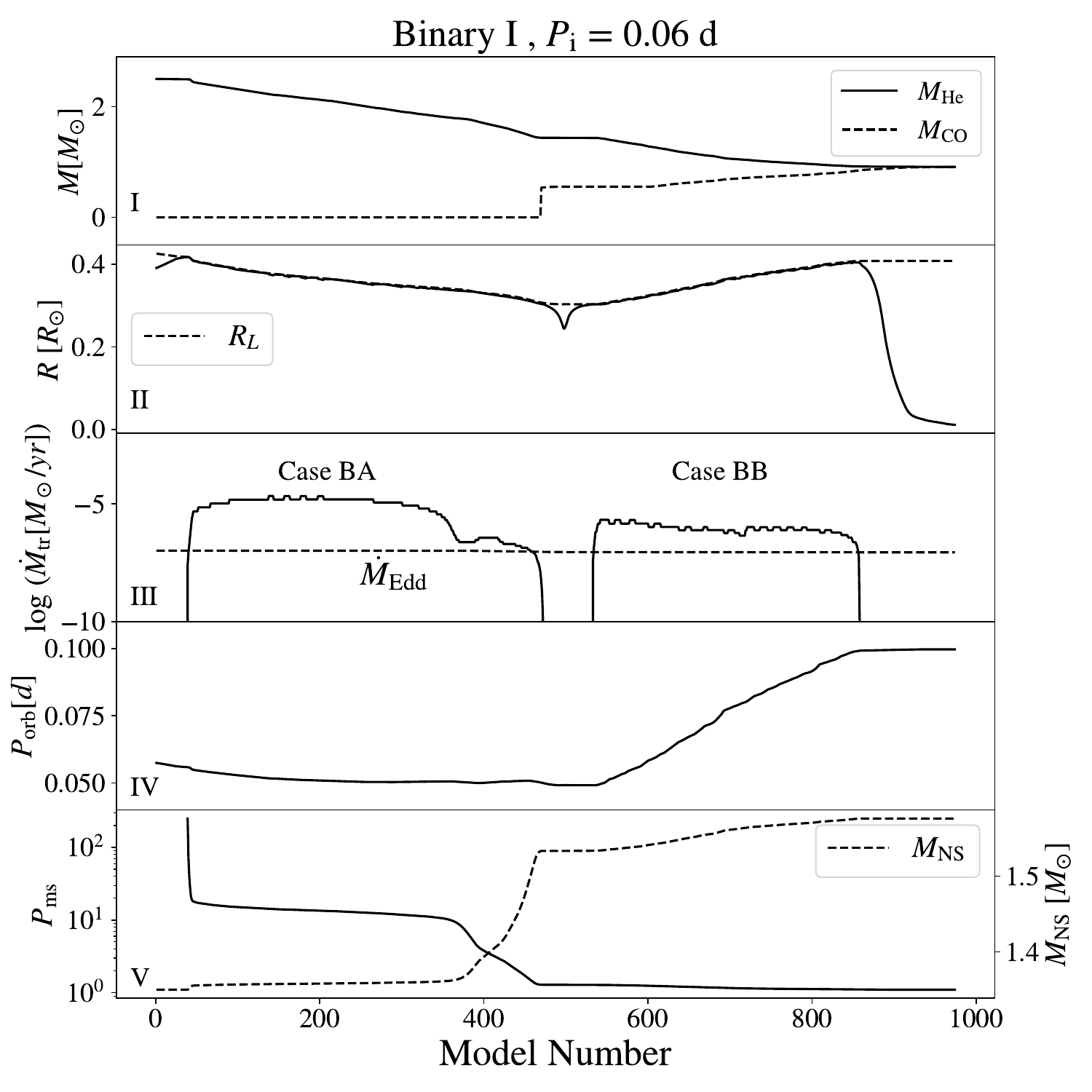}
     \caption{He-star/CO core mass (I), He-star radius/RLOF (II), mass transfer rate (III), orbital period (IV), and spin period (solid line)/mass (dashed line) (V) of the recycled NS as a function of model number for a $2.5\,M_\odot$ He star and NS companion. The initial orbital period is 0.06 d. The dashed line denotes the standard Eddington accretion rate.}
     \label{Binary_I} 
\end{figure}

In the following, we present the detailed evolutionary behavior of the three binary sequences. Figure~\ref{Binary_I} shows the evolution of the system with an initial orbital period of 0.06 d. During core helium burning, the He star gradually expands until it fills its Roche lobe, thereby initiating mass transfer onto the NS companion (Case BA).
Because the He star is initially more massive than the NS, mass transfer leads to a reduction in the orbital separation due to conservation of orbital angular momentum.

\begin{figure}[h]
     \centering
         \includegraphics[width=0.49\textwidth]{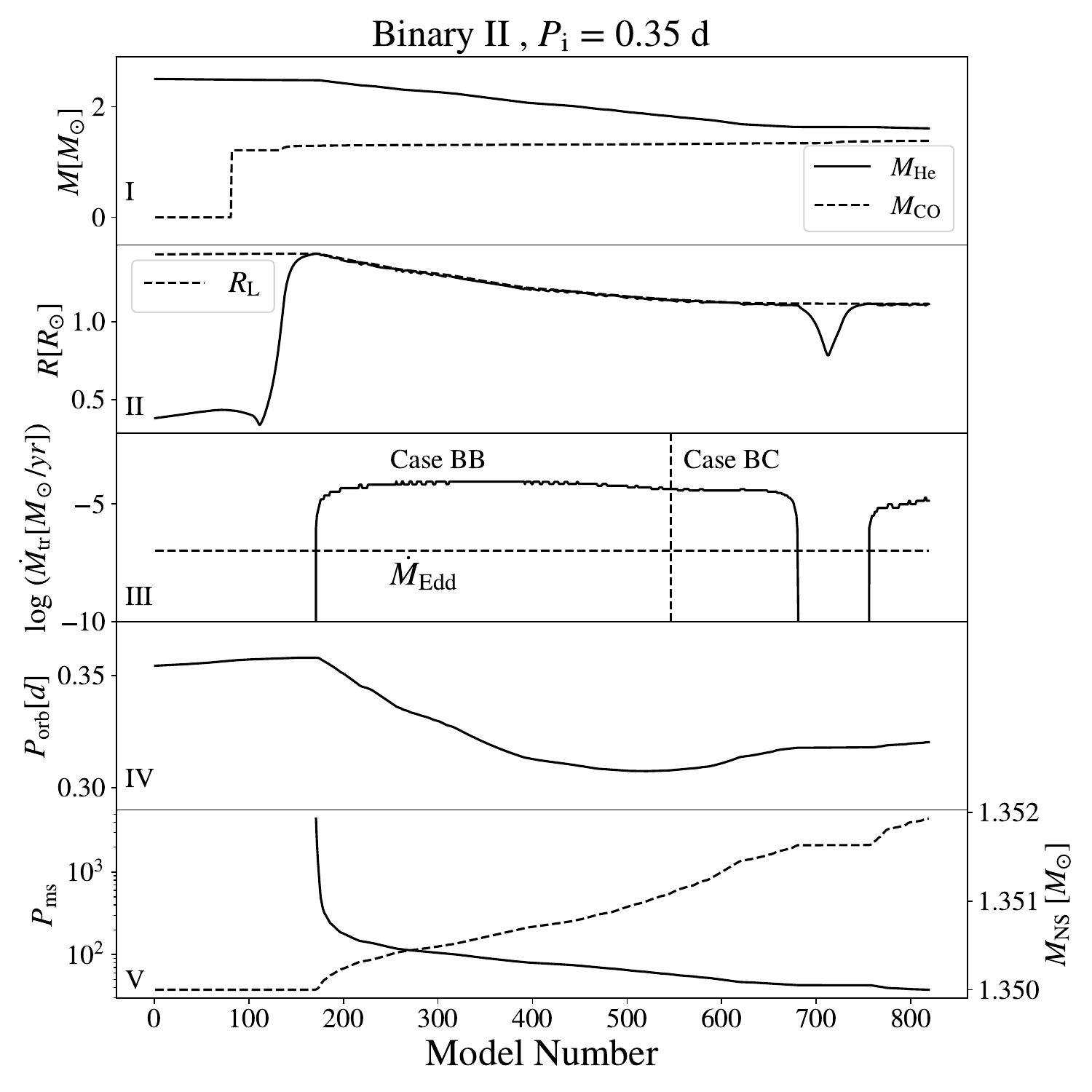}
     \caption{As in Figure~\ref{Binary_I}, but for the initial orbital period $P_i = 0.35$ d. The vertical dashed line marks the onset of the Case BC mass transfer.}
     \label{Binary_II} 
\end{figure}

\begin{figure}[h]
     \centering
             \includegraphics[width=0.49\textwidth]{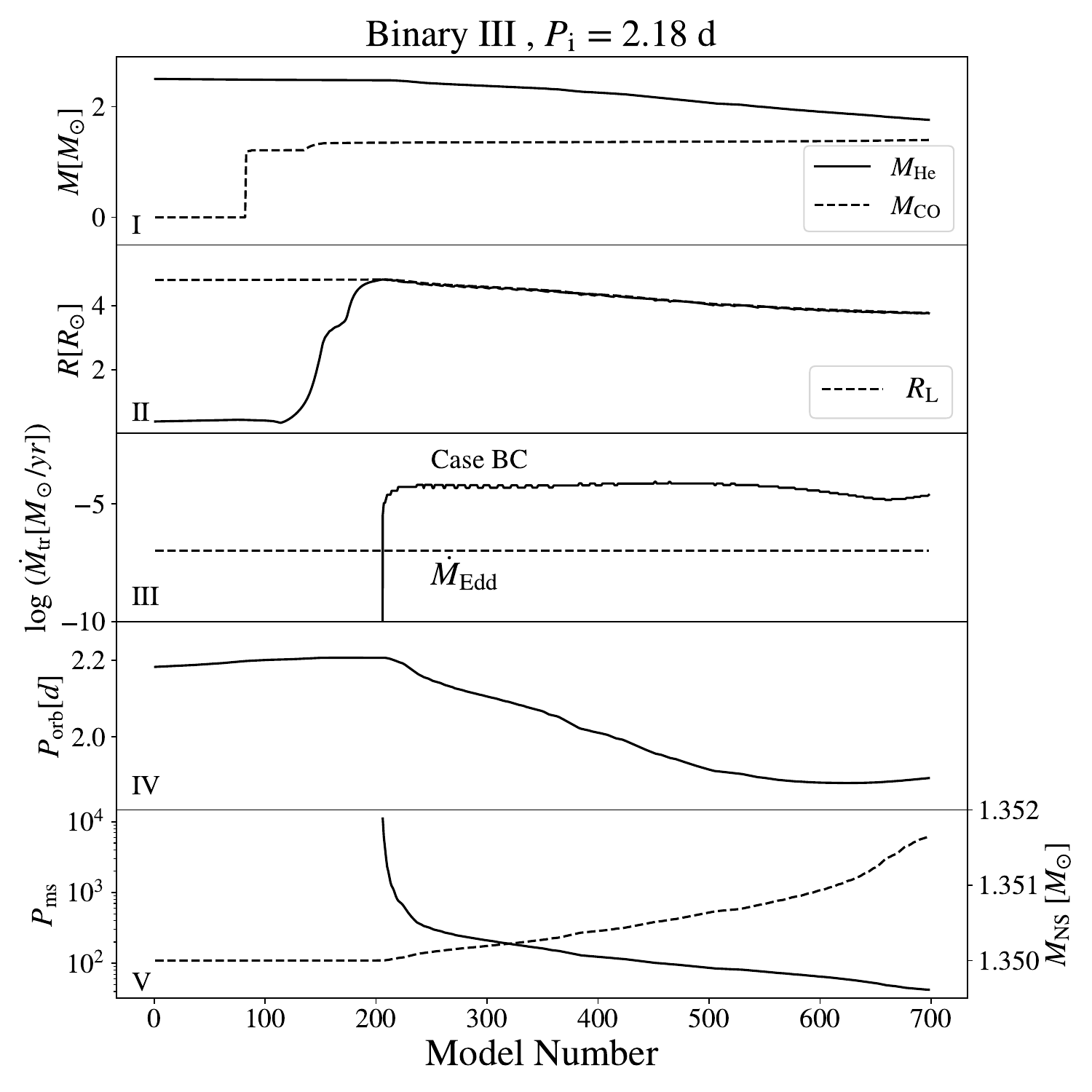}
         \caption{As in Figure~\ref{Binary_I}, but for the initial orbital period $P_i = 2.18$ d.}
     \label{Binary_III} 
\end{figure}

After central helium exhaustion, the He star contracts temporarily, causing the mass transfer to cease. As the core contracts and heats up, helium shell burning is ignited, leading to the expansion of the stellar envelope. This triggers a second phase of mass transfer (Case BB). By this stage, the He star has become less massive than the NS, and the orbit widens during the Case BB mass transfer. The system eventually detaches as the He star undergoes rapid contraction.

We terminate the evolution when the central Coulomb coupling parameter $\Gamma_C$ exceeds 10, which is taken to indicate the formation of a white dwarf (WD) \citep{Choi2016}. At the end of the calculation, the He star consists of a carbon–oxygen core of $\sim\, 0.91\, M_\odot$ surrounded by a thin helium shell of $\sim\,0.003\, M_\odot$. The NS has accreted a total mass of $\Delta M_{\rm acc} \sim 0.2259\, M_\odot$. Using the relation (see Eq.~\ref{ns_spin1}) between the NS spin period and the accreted mass from \cite{Tauris2012}, we estimate a spin period of $P_{\rm ms} \sim\, 1.1\, ms$.

\begin{equation}\label{ns_spin1}
P^{4/3}_{\rm ms} = 0.22 M_\odot \frac{(M_{\rm NS}/M_\odot)^{1/3}}{\Delta M_{\rm acc}}.
\end{equation}

Similar to the binary sequence discussed above, the detailed evolution of the other two systems is presented in Figure~\ref{Binary_II} ($P_{\rm i} = 0.35$ d; Binary II) and Figure~\ref{Binary_III} ($P_{\rm i} = 2.18$ d; Binary III). For the moderately wider initial orbital period of 0.35 d, mass transfer is initiated when the He star expands during the He-shell burning phase (Case BB) and continues into the C-core burning phase (Case BC). We note that the mass transfer is temporarily interrupted during the later phase (see Panel III). This occurs because the He star contracts after being deeply stripped in the earlier mass-transfer phase. Subsequently, the outer helium layers re-expand, driven primarily by energy transport from core carbon burning. We terminate the evolution at central carbon depletion. At this stage, the He star consists of a carbon–oxygen core of $\sim 1.38\,M_\odot$ surrounded by a thin helium shell of $\sim 0.224\,M_\odot$. Meanwhile, the NS accretes a total mass of $\Delta M_{\rm acc} \sim 0.0019\,M_\odot$, corresponding to an estimated spin period of $P_{\rm ms} \sim 37.6\,\mathrm{ms}$. 

For Binary III, with a wider initial orbital period of $P_{\rm i} = 2.18$ d, the evolution differs in several key aspects. Owing to the larger orbital separation, the onset of mass transfer is significantly delayed until the He star reaches a more advanced evolutionary stage and undergoes more substantial expansion. As a result, the mass-transfer phase is shorter-lived and proceeds less efficiently than in Binary II. At the stage of central carbon depletion, the He star in Binary III retains a more massive helium envelope than in Binary II (The He star consists of a carbon–oxygen core of $\sim 1.40\,M_\odot$ surrounded by a thin helium shell of $\sim 0.364\,M_\odot$), consistent with the reduced efficiency of mass removal. 

Meanwhile, the NS accretes slightly less material ($\Delta M_{\rm acc} \sim 0.0016\,M_\odot$), leading to a correspondingly longer spin period of $P_{\rm ms} \sim 42.4\, ms$. Overall, as the initial orbital period increases for the three binary systems, mass removal becomes progressively less efficient, leaving behind a more massive helium envelope, while the NS accretes less material and thus is spun up to lower rotational frequencies. For comparison, the $2.5\, M_\odot$ single He star produces a more massive carbon–oxygen core ($\sim 1.42\, M_\odot$) and retains a larger helium-rich envelope ($\sim 1.059\, M_\odot$) at the same evolutionary stage, since no additional mass loss occurs via binary interaction.

\subsubsection{Newly formed neutron star} 
\begin{figure}[h]
     \centering
         \includegraphics[width=0.49\textwidth]{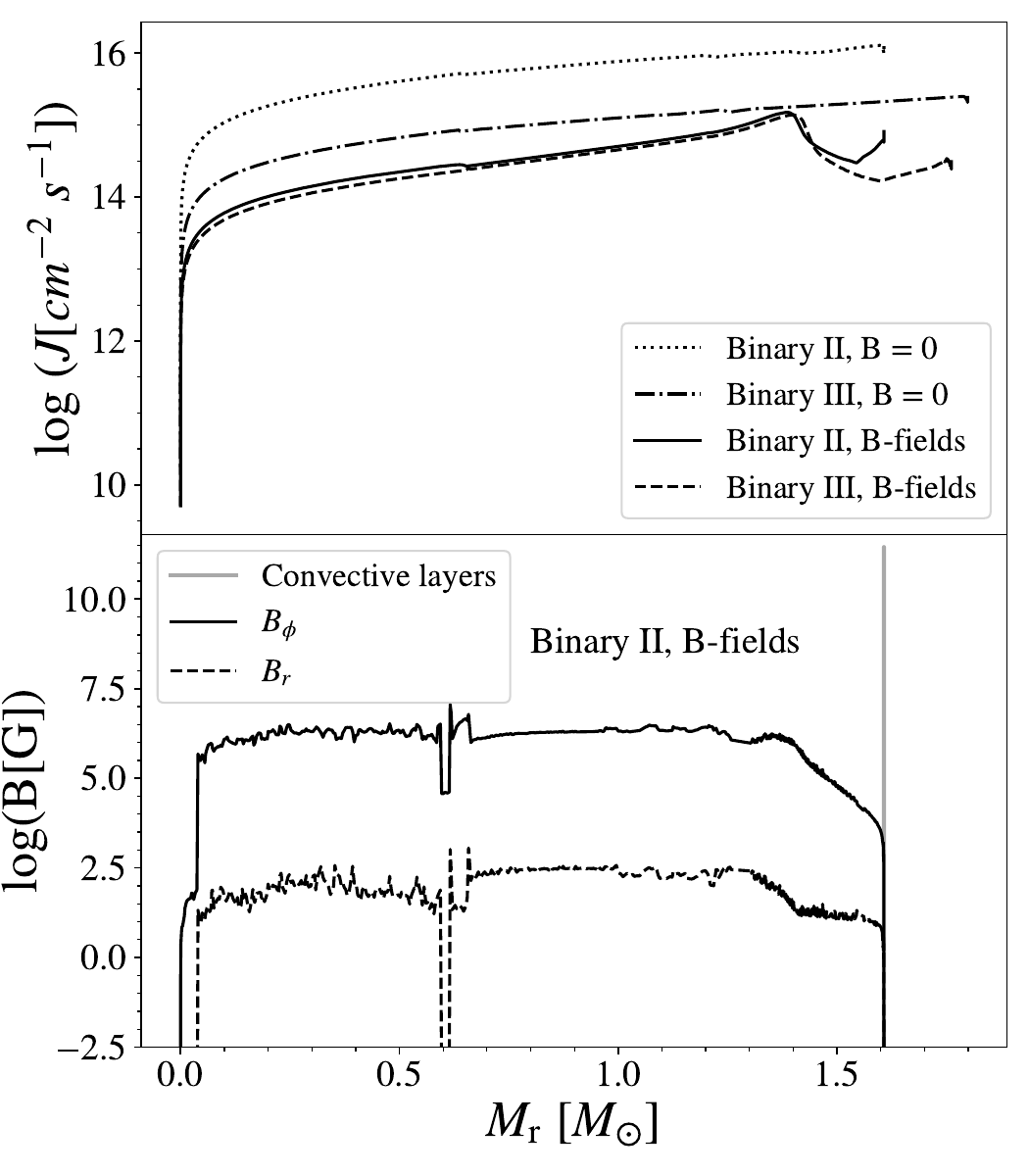}
     \caption{Upper panel: specific angular momentum distribution for the models of Binary II (dotted line: B = 0, solid line: B-fields) and Binary III (dash-dotted line: B = 0, dashed line: B-fields). Bottom panel: the azimuthal and radial components of the magnetic field ($B_\phi$ and $B_r$) of Binary II (B-fields). The vertical lines represent the region of the convective layers.}
     \label{B_J} 
\end{figure}

As the resulting carbon–oxygen core masses in Binary II and Binary III fall within the ECSN range ($1.37\, M_\odot \lesssim M_{\rm CO} \lesssim 1.43\, M_\odot$; see the discussion in \citealt{Tauris2015}), we estimate the NS gravitational mass to be $1.26\, M_\odot$ using the prescription of \cite{Fryer2012}. Following the same approach as in \cite{Qin2024}, we further estimate the NS’s initial rotational energy. To compute the moment of inertia of the NS, we adopt the empirical relation from \citet{Lattimer2001}:

\begin{equation}\label{ns_spin}
I_{\rm NS} = 0.35 M_{\rm NS} R_{\rm NS}^2,
\end{equation}

\noindent
where $M_{\rm NS}$ and $R_{\rm NS}$ denote the NS mass ($M_{\rm NS} = 1.26\, M_\odot$) and radius ($R_{\rm NS} = 12.5\, km$) \citep{gw190425,Landry2020}, respectively. Using the above relations, we obtain the initial rotational energy of the NS, $E_{\rm rot} \sim\, 2.4 \times 10^{50}$ erg for Binary II and $E_{\rm rot} \sim\, 1.8 \times 10^{50}$ erg for Binary III. We then compare the resulting specific angular momentum distribution of the He-star models at the central carbon depletion, accounting for the effects of the Spruit–Tayler dynamo \citep{Spruit2002}, which operates in radiative regions where differential rotation amplifies a weak seed radial magnetic field into a strong toroidal component via shear \cite[e.g.,][]{Spruit2002,Heger2005,Petrovic2005}. As shown in the top panel of Figure~\ref{B_J}, the inclusion of the Spruit–Tayler dynamo reduces the angular momentum content of the He-star models by approximately one to two orders of magnitude. For comparison, neglecting the Spruit–Tayler dynamo yields significantly larger NS rotational energies, i.e., $E_{\rm rot} \sim 5.7 \times 10^{52}$ erg for Binary II and $E_{\rm rot} \sim 1.5 \times 10^{51}$ erg for Binary III. In the case of Binary II with the Spruit–Tayler dynamo included, the He star becomes almost fully radiative, with only very thin convective regions. In the bottom panel, we present the distributions of the azimuthal ($B_\phi$) and radial ($B_r$) magnetic field components. As expected, magnetic fields are generated predominantly in radiative regions of the star, consistent with the Spruit–Tayler dynamo mechanism \citep{Spruit2002}.

Following the same method as in \cite{Song2023}, we evaluate the average magnetic fields within the carbon–oxygen core, defined as

\begin{equation}
    \langle B_\phi \rangle = \frac{\int_{0}^{M_{\rm CO}} B_\phi(m) \, dm}{\int_{0}^{M_{\rm CO}} dm}.
    \label{eq:avg_B_phi}
\end{equation}

Assuming magnetic flux conservation during the collapse of the carbon–oxygen core, we estimate the resulting NS magnetic field strength to be $\sim 6.9 \times 10^{12} G$ for Binary II and $\sim 4.1 \times 10^{12} G$ for Binary III. We further estimate the natal spin periods of the NSs by using the angular momentum contained in the carbon–oxygen core\footnote{We note that the NS spin can be approximated using a power-law relationship with the pre-explosion mass and orbital period \cite[see more details in][]{Fuller2022}.},
\begin{equation}
    P_{\rm NS}=\frac{2\pi I_{\rm NS}}{J_{\rm CO}}.
\end{equation}
The resulting spin periods are $P_{\rm NS} \sim 7.9\, ms$ for Binary II and $\sim 9.1\, ms$ for Binary III. For comparison, if the Spruit–Tayler dynamo is not included, the predicted spin periods are significantly shorter, i.e., $P_{\rm NS} \sim 0.6\, ms$ for Binary II and $\sim 3.5\, ms$ for Binary III, reflecting the substantially larger retained core angular momentum in the absence of magnetic angular-momentum transport.

\subsection{Parameter space analysis}
In this section, we further investigate the impact of close binary interactions on the evolution of low-mass He stars across a broad parameter space. To this end, we consider binary systems with initial orbital periods ranging from 0.04 to 40 d and initial He-star masses of $\sim 2.4 - 2.7\, M_\odot$ at solar metallicity, corresponding to the mass range in which ECSNe are expected to occur. In addition, we include low-metallicity models with $Z = 0.01\, Z_\odot$ to explore the metallicity dependence of the results. Unless otherwise stated, all He-star models incorporate the effects of the Spruit–Tayler dynamo.

\begin{figure*}[h]
     \centering
     \includegraphics[width=0.9\textwidth]{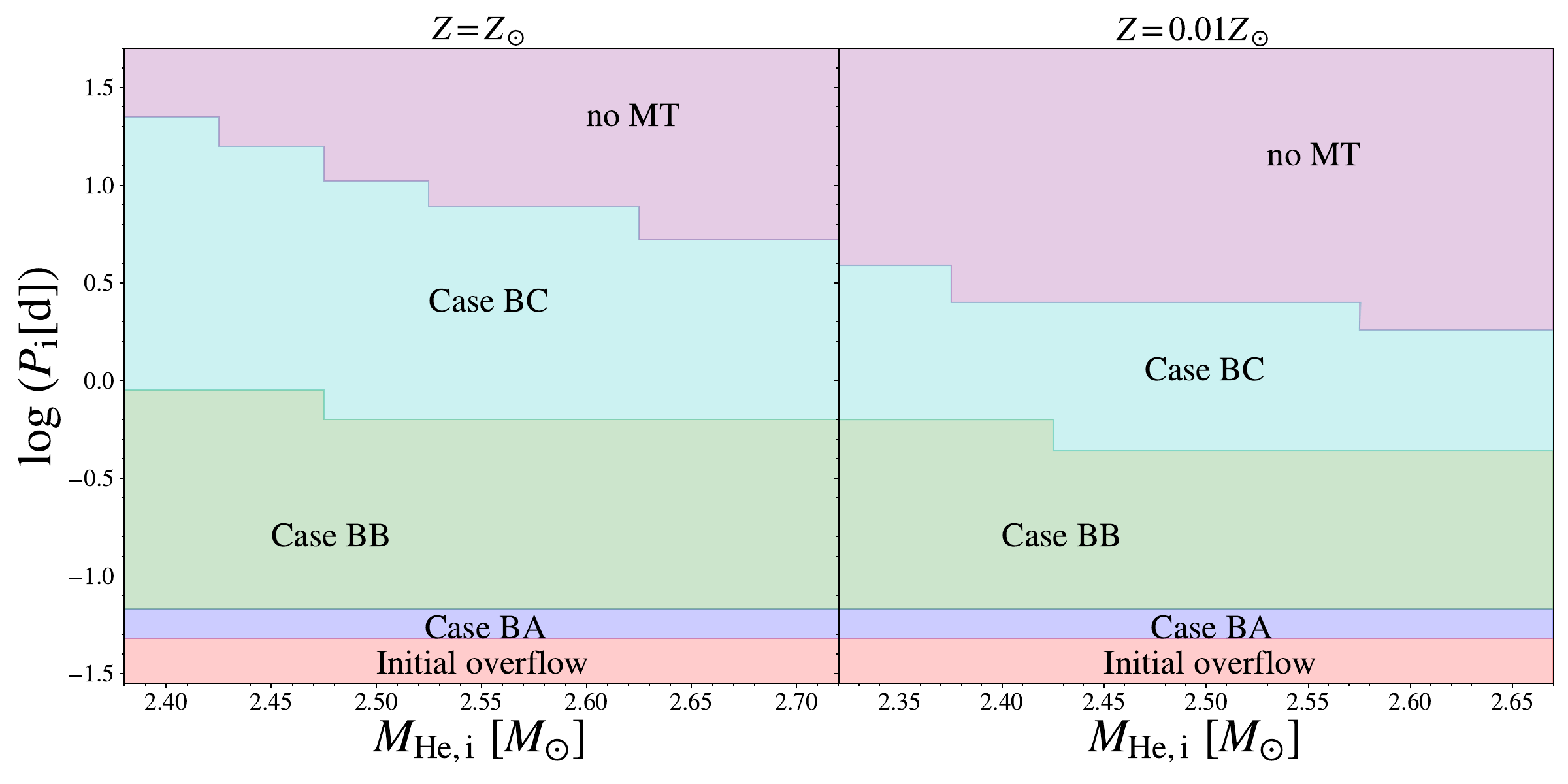}
     \caption{Outcome of binary interactions consisting of a He star and a 1.35 $M_\odot$ NS companion as a function of the initial values of orbital period and He star mass. The colored regions represent the parameter spaces for various mass transfer phases, i.e., light red: Initial overflow; light blue: Case BA; light green: Case BB; light cyan: Case BC; light purple: No mass transfer (No MT). Left panel: high metallicity ($Z = Z_{\odot}$); right panel: low metallicity ($Z = 0.01 Z_{\odot}$).}
     \label{mt} 
\end{figure*}

The left panel of Figure~\ref{mt} illustrates the different mass-transfer cases as functions of the initial orbital period and He-star mass at solar metallicity.
Systems with sufficiently wide initial orbits avoid mass transfer throughout their evolution. Since more massive He stars generally expand to smaller radii, binaries with higher initial He-star mass initiate mass transfer at narrower orbital separations. Additionally, systems with shorter initial orbital periods experience Roche-lobe overflow at earlier evolutionary stages, reflecting the gradual radial expansion of low-mass He stars during their evolution (see Figure~\ref{HR}). Binaries with extremely short initial orbital periods undergo Roche-lobe overflow almost immediately after the start of the evolution ``initial overflow'' models) and are therefore expected to merge; such systems are excluded from further discussion. A similar behavior is found for the low-metallicity models shown in the right panel. In particular, the upper orbital-period boundaries for both Case BB and Case BC mass transfer shift toward shorter periods, because low-metallicity He stars remain more compact during their evolution.

\begin{figure*}[h]
     \centering
     \includegraphics[width=0.9\textwidth]{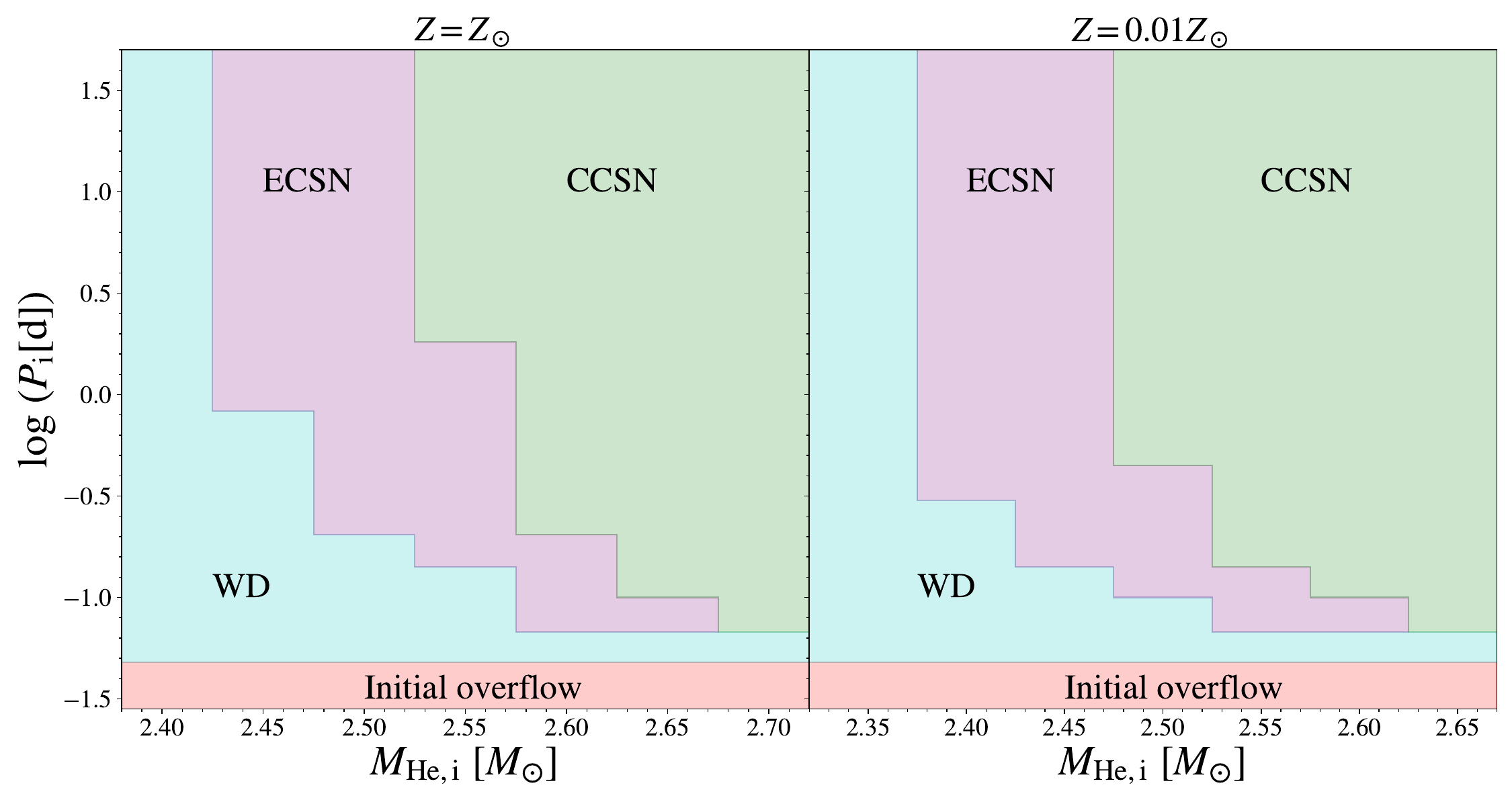}
     \caption{As in Figure~\ref{mt}, but the colored regions represent different evolutionary fates of He stars, i.e., light cyan: White dwarf (WD); light purple: Electron-capture supernovae (ECSN); light green: Core-collapse supernova (CCSN).}
     \label{fate} 
\end{figure*}

Figure~\ref{fate} presents the various evolutionary outcomes of He stars for different initial conditions. As shown in the left panel, ECSNe occur within a relatively narrow initial He-star mass range of $2.42 - 2.67\, M_\odot$. For systems with shorter initial orbital periods, mass transfer is initiated earlier and removes the He-star envelope more efficiently. Consequently, a larger initial He-star mass is required to maintain a sufficiently massive core for ECSN. At lower metallicity, stellar winds are weaker, enabling He stars to retain more mass throughout their evolution. As a result, the ECSN progenitor mass range shifts toward lower initial masses, namely $2.37 - 2.62\, M_\odot$, as shown in the right panel.

Within the explored parameter space, the resulting ECSNe produce NSs with spin periods ranging from $83.8$ to $7.7\, ms$ at solar metallicity, and from $81.5$ to $7.7\, ms$ at $0.01\, Z_\odot$. Correspondingly, the rotational energies in the range of $E_{\rm rot} \sim 2.6 \times 10^{48}$ – $2.5 \times 10^{50} \mathrm{erg}$ at solar metallicity and $2.7 \times 10^{48}$ – $2.6 \times 10^{50} \mathrm{erg}$ at $0.01\, Z_\odot$. Additionally, the recycling NSs are predicted to have spin periods from $662.6$ to $14.9\, ms$ at solar metallicity, and from $201.1$ to $15.9\, ms$ at $0.01\, Z_\odot$.

He-rich stars that undergo Case BB/BC mass transfer and experience further envelope stripping are generally expected to produce ultra-stripped supernovae (ultra-stripped SNe; \citealt{Tauris2015}). In our ECSN models, the helium-envelope masses at carbon depletion range from 0.14 $M_\odot$ to 1.1 $M_\odot$, which are somewhat lower than those of CCSN progenitors \citep{Qin_gw190425}. Spectral modeling by \citet{Hachinger2012} suggests that at least $\sim 0.06\, M_\odot$ of helium is required for helium lines to be detectable in the supernova spectrum; below this threshold, the event is likely to be classified as a Type Ic supernova. Since all of our ECSN models retain helium-envelope masses above this limit, they are expected to produce Type Ib SNe. However, as demonstrated by \citet{Wu2022}, He stars re-expand during the oxygen/neon-burning phase, which can trigger mass transfer and further strip the remaining helium envelope. Similarly, \citet{Guo2024} showed that continued late-stage envelope stripping may reduce the retained helium mass sufficiently for some progenitors to explode as Type Ic supernovae. We also find that the rotational energies of the neutron stars produced by ECSNe in our models are approximately two orders of magnitude lower than those predicted for Fe CCSNe \citep{Qin_gw190425}.

\subsection{Connection to the observed DNS in the Milky Way}
The kick velocity imparted onto the newborn NS can alter the post-SN orbital separation and eccentricity of the binary system. NSs formed through ECSNe are generally expected to receive relatively small natal kicks because of their low explosion energies \cite[$V_{\rm k}$ $\leq$ 50 $\rm km$ $\rm s^{-1}$, e.g.,][for a recent review and references therein]{Wang2026}. Furthermore, ECSNe occurring in ultra-stripped supernovae are associated with extremely low ejecta masses, which are also expected to produce small kick velocities \cite[see discussions in][]{Tauris2015}.

\begin{figure*}[h]
    \centering
        \includegraphics[width=0.45\textwidth]{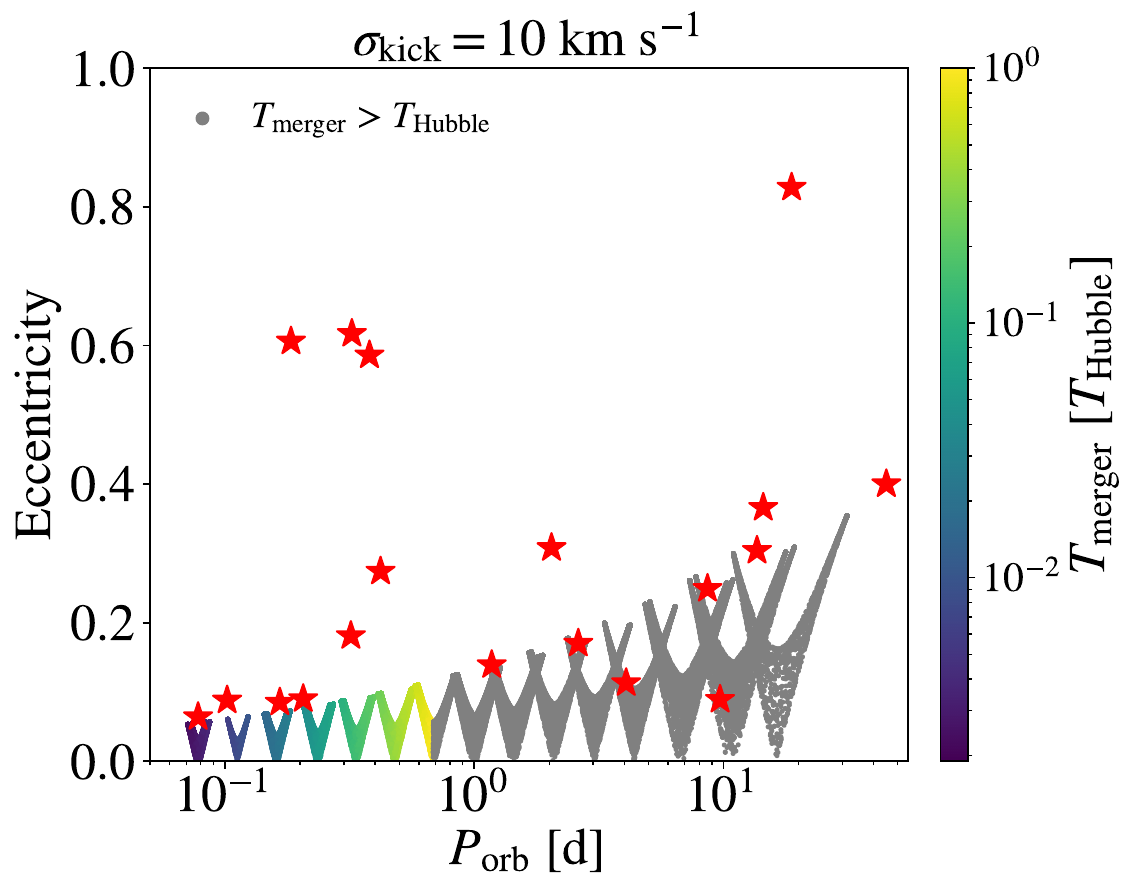}
        \includegraphics[width=0.45\textwidth]{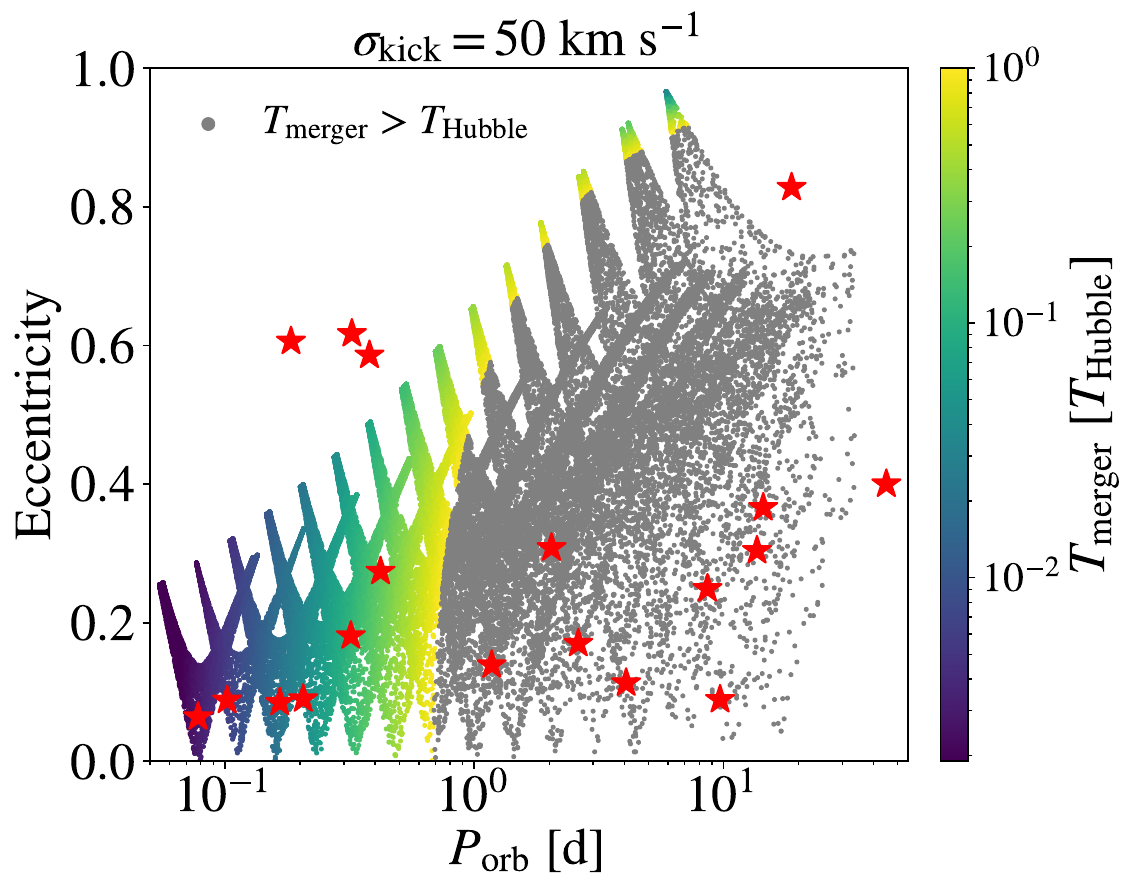}
        \caption{Ratio of the merger time of the DNS to the Hubble time as a function of eccentricity and orbital period (pre-SN orbital period on the X axis). Left panel: $\sigma_{\rm ECSN} = 10\, \rm km\, s^{-1}$; Right panel: $\sigma_{\rm ECSN} = 50\, \rm km\, s^{-1}$. Simulated data in gray are for the systems whose merger time is longer than the Hubble time. The red stars denote the observed Galactic DNS systems thanks to the ATNF Pulsar Catalogue \citep{Manchester2005}.}
        \label{kick} 
\end{figure*}

We follow the same methodology as in \cite{Hu2022} to simulate SN kicks. Specifically, natal kicks are randomly sampled 2000 times from a Maxwellian distribution with dispersions of $\sigma_{\rm ECSN} =$ 10, 50 $\rm km$ $\rm s^{-1}$. Rather than assuming fixed orbital periods (i.e., 0.1, 1.0, 10, and 40 d) as adopted by \citet{Guo2024}, we derive the post-SN binary properties from the pre-SN parameters of systems located within the ECSN formation parameter space (see the left panel of Figure~\ref{fate}). In general, the eccentricity and orbital period of the NS binary can be significantly altered after the He star undergoes a SN explosion. The orientation of the natal kick velocity is characterized by two angles: the angle between the kick velocity $V_k$ and the pre-SN orbital velocity ($0^\circ \le \theta \le 180^\circ$), and the azimuthal angle describing the direction of $V_k$ out of the orbital plane ($-180^\circ < \phi \le 180^\circ$). Figure~\ref{kick} shows the eccentricities and orbital periods of the post-SN DNS systems. Consistent with the results of \citet{Guo2024}, we also identify a ``V''-shaped correlation between the eccentricity and orbital period of post-SN binaries. As expected, allowing larger natal kick velocities enables the formation of more observed DNS systems with high eccentricities. 

Following the formation of the newborn NS through ECSN, we calculate the merger timescales due to gravitational-wave emission using the formalism of \cite{Peters1964}. As shown in Figure~\ref{kick}, observed DNS systems with orbital periods longer than $\sim 1.0\,\mathrm{d}$ are unlikely to merge within a Hubble time and therefore are not expected to contribute significantly to the LIGO–Virgo–KAGRA gravitational-wave sources.

\section{Conclusions and discussion}\label{sect4}
In this study, we adopt the wind prescription of V2017 to investigate the evolution of low-mass He stars ($2.5 - 5\, M_\odot$) at different metallicities. 
We find that rotation has only a modest impact on the overall evolution of low-mass He stars. In addition, we perform detailed binary evolution calculations for low-mass He stars in close binary systems. The evolutionary outcomes strongly depend sensitively on the initial orbital period. Systems with shorter initial orbital periods experience Roche-lobe overflow at earlier evolutionary stages. When the companion is a NS, accretion during the mass-transfer phase can efficiently spin up the NS, potentially producing millisecond pulsars.

To systematically explore the effects of rotation, mass transfer, and tidal interaction, we conduct a parameter-space study of low-mass He stars in close binaries. We find that ECSNe are produced within a relatively narrow range of initial He-star masses of $2.42$ – $2.67\, M_\odot$ at $Z_\odot$ and $2.37$ – $2.62\, M_\odot$ for 0.01 $Z_\odot$. The neutron stars formed through ECSNe in our models have spin periods ranging from $83.8$ to $7.7\, ms$ at solar metallicity, and from $81.5$ to $7.7\, ms$ at $0.01\, Z_\odot$. The estimated magnetic-field strengths are of the order of $10^{12}$ G, approximately two orders of magnitude lower than those of Fe core-collapse supernovae \citep{Song2023}. The corresponding rotational energies vary
in a range of $E_{\rm rot} \sim 2.6 \times 10^{48}$ – $2.5 \times 10^{50} \mathrm{erg}$ at $Z_\odot$ and $2.7 \times 10^{48}$ – $2.6 \times 10^{50} \mathrm{erg}$ at $0.01\, Z_\odot$, which are approximately two orders of magnitude lower than those predicted for Fe CCSNe \citep{Qin_gw190425}. This suggests that ECSNe provide a substantially smaller rotational energy reservoir, although these energies would be further reduced if efficient angular-momentum transport mechanisms, such as the Spruit–Tayler dynamo, were included.

Given the weak wind mass-loss prescription adopted in our models and the termination of our calculations at central carbon depletion, the progenitors retain relatively massive helium envelopes, suggesting that they are likely to produce Type Ib supernovae. However, the subsequent late-stage evolution may significantly modify the remaining helium content. In particular, helium stars can re-expand during oxygen/neon burning and undergo additional episodes of mass transfer \citep{Wu2022}, while continued envelope stripping in the late evolutionary stages may further reduce the helium mass \citep{Guo2024}. Therefore, some ECSN progenitors may lose sufficient helium to produce Type Ic supernovae before core collapse.

Finally, by comparing our models with observed Galactic double DNS systems, we find that most systems can be reproduced in the eccentricity–orbital-period plane when relatively high natal kick velocities (e.g., 50 $\rm km$ $\rm s^{-1}$) are adopted, consistent with the findings of \citet{Guo2024}. Our results further indicate that the observed double neutron star systems with orbital periods longer than $\sim 1.0\,\mathrm{d}$ are unlikely to merge within a Hubble time and therefore are not expected to contribute significantly to the population of gravitational-wave sources detectable by the LIGO–Virgo–KAGRA observatories.

\begin{acknowledgements}
We thank Hai-Liang Chen, Tao Wu, and Rui-Chong Hu for helpful comments on the manuscript. Y.Q. acknowledges support from the National Natural Science Foundation of China (grant Nos. 12473036 and 12573045). This work was partially supported by the Jiangxi Provincial Natural Science Foundation (grant Nos. 20242BAB26012 and 20224ACB211001) and by Anhui Province Graduate Education Quality Engineering Project (grant No. 2024qyw/sysfkc012). G.M. has received funding from the European Research Council (ERC) under the European Union’s Horizon 2020 research and innovation program (grant agreement No 833925, project STAREX). H.F. Song is supported by the National Natural Science Foundation of China (grant Nos. 12173010 and 12573034). All figures are made with the free Python module Matplotlib \citep{Hunter2007}.
\end{acknowledgements}

\bibliography{ref}

@ARTICLE{Tauris2017,
       author = {{Tauris}, T.~M. and {Kramer}, M. and {Freire}, P.~C.~C. and {Wex}, N. and {Janka}, H.-T. and {Langer}, N. and {Podsiadlowski}, Ph. and {Bozzo}, E. and {Chaty}, S. and {Kruckow}, M.~U. and {van den Heuvel}, E.~P.~J. and {Antoniadis}, J. and {Breton}, R.~P. and {Champion}, D.~J.},
        title = "{Formation of Double Neutron Star Systems}",
      journal = {\apj},
         year = 2017,
        month = sep,
       volume = {846},
       number = {2},
          eid = {170},
        pages = {170},
          doi = {10.3847/1538-4357/aa7e89},
archivePrefix = {arXiv},
       eprint = {1706.09438},
 primaryClass = {astro-ph.HE},
       adsurl = {https://ui.adsabs.harvard.edu/abs/2017ApJ...846..170T}
}

@article{Paxton2011,
       author = {{Paxton}, Bill and {Bildsten}, Lars and {Dotter}, Aaron and {Herwig}, Falk and {Lesaffre}, Pierre and {Timmes}, Frank},
        title = "{Modules for Experiments in Stellar Astrophysics (MESA)}",
      journal = {\apjs},
         year = 2011,
        month = jan,
       volume = {192},
       number = {1},
          eid = {3},
        pages = {3},
          doi = {10.1088/0067-0049/192/1/3},
archivePrefix = {arXiv},
       eprint = {1009.1622},
 primaryClass = {astro-ph.SR},
       adsurl = {https://ui.adsabs.harvard.edu/abs/2011ApJS..192....3P}
}

@article{Paxton2013,
       author = {{Paxton}, Bill and {Cantiello}, Matteo and {Arras}, Phil and {Bildsten}, Lars and {Brown}, Edward F. and {Dotter}, Aaron and {Mankovich}, Christopher and {Montgomery}, M.~H. and {Stello}, Dennis and {Timmes}, F.~X. and {Townsend}, Richard},
        title = "{Modules for Experiments in Stellar Astrophysics (MESA): Planets, Oscillations, Rotation, and Massive Stars}",
      journal = {\apjs},
         year = 2013,
        month = sep,
       volume = {208},
       number = {1},
          eid = {4},
        pages = {4},
          doi = {10.1088/0067-0049/208/1/4},
archivePrefix = {arXiv},
       eprint = {1301.0319},
 primaryClass = {astro-ph.SR},
       adsurl = {https://ui.adsabs.harvard.edu/abs/2013ApJS..208....4P}
}

@ARTICLE{Paxton2015,
       author = {{Paxton}, Bill and {Marchant}, Pablo and {Schwab}, Josiah and {Bauer}, Evan B. and {Bildsten}, Lars and {Cantiello}, Matteo and {Dessart}, Luc and {Farmer}, R. and {Hu}, H. and {Langer}, N. and {Townsend}, R.~H.~D. and {Townsley}, Dean M. and {Timmes}, F.~X.},
        title = "{Modules for Experiments in Stellar Astrophysics (MESA): Binaries, Pulsations, and Explosions}",
      journal = {\apjs},
         year = 2015,
        month = sep,
       volume = {220},
       number = {1},
          eid = {15},
        pages = {15},
          doi = {10.1088/0067-0049/220/1/15},
archivePrefix = {arXiv},
       eprint = {1506.03146},
 primaryClass = {astro-ph.SR},
       adsurl = {https://ui.adsabs.harvard.edu/abs/2015ApJS..220...15P}
}

@ARTICLE{Paxton2018,
       author = {{Paxton}, Bill and {Schwab}, Josiah and {Bauer}, Evan B. and {Bildsten}, Lars and {Blinnikov}, Sergei and {Duffell}, Paul and {Farmer}, R. and {Goldberg}, Jared A. and {Marchant}, Pablo and {Sorokina}, Elena and {Thoul}, Anne and {Townsend}, Richard H.~D. and {Timmes}, F.~X.},
        title = "{Modules for Experiments in Stellar Astrophysics (MESA): Convective Boundaries, Element Diffusion, and Massive Star Explosions}",
      journal = {\apjs},
         year = 2018,
        month = feb,
       volume = {234},
       number = {2},
          eid = {34},
        pages = {34},
          doi = {10.3847/1538-4365/aaa5a8},
archivePrefix = {arXiv},
       eprint = {1710.08424},
 primaryClass = {astro-ph.SR},
       adsurl = {https://ui.adsabs.harvard.edu/abs/2018ApJS..234...34P}
}

@ARTICLE{Paxton2019,
       author = {{Paxton}, Bill and {Smolec}, R. and {Schwab}, Josiah and {Gautschy}, A. and {Bildsten}, Lars and {Cantiello}, Matteo and {Dotter}, Aaron and {Farmer}, R. and {Goldberg}, Jared A. and {Jermyn}, Adam S. and {Kanbur}, S.~M. and {Marchant}, Pablo and {Thoul}, Anne and {Townsend}, Richard H.~D. and {Wolf}, William M. and {Zhang}, Michael and {Timmes}, F.~X.},
        title = "{Modules for Experiments in Stellar Astrophysics (MESA): Pulsating Variable Stars, Rotation, Convective Boundaries, and Energy Conservation}",
      journal = {\apjs},
         year = 2019,
        month = jul,
       volume = {243},
       number = {1},
          eid = {10},
        pages = {10},
          doi = {10.3847/1538-4365/ab2241},
archivePrefix = {arXiv},
       eprint = {1903.01426},
 primaryClass = {astro-ph.SR},
       adsurl = {https://ui.adsabs.harvard.edu/abs/2019ApJS..243...10P}
}

@ARTICLE{Jermyn2023,
       author = {{Jermyn}, Adam S. and {Bauer}, Evan B. and {Schwab}, Josiah and {Farmer}, R. and {Ball}, Warrick H. and {Bellinger}, Earl P. and {Dotter}, Aaron and {Joyce}, Meridith and {Marchant}, Pablo and {Mombarg}, Joey S.~G. and {Wolf}, William M. and {Sunny Wong}, Tin Long and {Cinquegrana}, Giulia C. and {Farrell}, Eoin and {Smolec}, R. and {Thoul}, Anne and {Cantiello}, Matteo and {Herwig}, Falk and {Toloza}, Odette and {Bildsten}, Lars and {Townsend}, Richard H.~D. and {Timmes}, F.~X.},
        title = "{Modules for Experiments in Stellar Astrophysics (MESA): Time-dependent Convection, Energy Conservation, Automatic Differentiation, and Infrastructure}",
      journal = {\apjs},
         year = 2023,
        month = mar,
       volume = {265},
       number = {1},
          eid = {15},
        pages = {15},
          doi = {10.3847/1538-4365/acae8d},
archivePrefix = {arXiv},
       eprint = {2208.03651},
 primaryClass = {astro-ph.SR},
       adsurl = {https://ui.adsabs.harvard.edu/abs/2023ApJS..265...15J}
}

@ARTICLE{Fragos2023,
       author = {{Fragos}, Tassos and {Andrews}, Jeff J. and {Bavera}, Simone S. and {Berry}, Christopher P.~L. and {Coughlin}, Scott and {Dotter}, Aaron and {Giri}, Prabin and {Kalogera}, Vicky and {Katsaggelos}, Aggelos and {Kovlakas}, Konstantinos and {Lalvani}, Shamal and {Misra}, Devina and {Srivastava}, Philipp M. and {Qin}, Ying and {Rocha}, Kyle A. and {Rom{\'a}n-Garza}, Jaime and {Serra}, Juan Gabriel and {Stahle}, Petter and {Sun}, Meng and {Teng}, Xu and {Trajcevski}, Goce and {Tran}, Nam Hai and {Xing}, Zepei and {Zapartas}, Emmanouil and {Zevin}, Michael},
        title = "{POSYDON: A General-purpose Population Synthesis Code with Detailed Binary-evolution Simulations}",
      journal = {\apjs},
         year = 2023,
        month = feb,
       volume = {264},
       number = {2},
          eid = {45},
        pages = {45},
          doi = {10.3847/1538-4365/ac90c1},
archivePrefix = {arXiv},
       eprint = {2202.05892},
 primaryClass = {astro-ph.SR},
       adsurl = {https://ui.adsabs.harvard.edu/abs/2023ApJS..264...45F}
}

@ARTICLE{lv2023,
       author = {{Lyu}, F. and {Yuan}, L. and {Wu}, D.~H. and {Guo}, W.~H. and {Wang}, Y.~Z. and {Yi}, S.~X. and {Tang}, Q.~W. and {Hu}, R. -C. and {Zhu}, J. -P. and {Shu}, X.~W. and {Qin}, Y. and {Liang}, E.~W.},
        title = "{Revisiting the properties of GW190814 and its formation history}",
      journal = {\mnras},
         year = 2023,
        month = nov,
       volume = {525},
       number = {3},
        pages = {4321-4328},
          doi = {10.1093/mnras/stad2538},
archivePrefix = {arXiv},
       eprint = {2308.09893},
 primaryClass = {astro-ph.HE},
       adsurl = {https://ui.adsabs.harvard.edu/abs/2023MNRAS.525.4321L}
}

@ARTICLE{Qin2024_gap,
       author = {{Qin}, Ying and {Wang}, Zhen-Han-Tao and {Meynet}, Georges and {Hu}, Rui-Chong and {Fu}, Chengjie and {Shu}, Xin-Wen and {Wang}, Zi-Yuan and {Yi}, Shuang-Xi and {Tang}, Qing-Wen and {Song}, Han-Feng and {Liang}, En-Wei},
        title = "{Origin of the black hole spin in lower-mass-gap black hole-neutron star binaries}",
      journal = {\aap},
         year = 2024,
        month = nov,
       volume = {691},
          eid = {L19},
        pages = {L19},
          doi = {10.1051/0004-6361/202452335},
archivePrefix = {arXiv},
       eprint = {2409.14476},
 primaryClass = {astro-ph.HE},
       adsurl = {https://ui.adsabs.harvard.edu/abs/2024A&A...691L..19Q}
}

@ARTICLE{Qin2023,
       author = {{Qin}, Y. and {Hu}, R.-C. and {Meynet}, G. and {Wang}, Y.~Z. and {Zhu}, J.-P. and {Song}, H.~F. and {Shu}, X.~W. and {Wu}, S.~C.},
        title = "{Merging binary black holes formed through double-core evolution}",
      journal = {\aap},
         year = 2023,
        month = mar,
       volume = {671},
          eid = {A62},
        pages = {A62},
          doi = {10.1051/0004-6361/202244712},
archivePrefix = {arXiv},
       eprint = {2301.04918},
 primaryClass = {astro-ph.HE},
       adsurl = {https://ui.adsabs.harvard.edu/abs/2023A&A...671A..62Q}
}

@ARTICLE{Qin2024,
       author = {{Qin}, Ying and {Zhu}, Jin-Ping and {Meynet}, Georges and {Zhang}, Bing and {Wang}, Fa-Yin and {Shu}, Xin-Wen and {Song}, Han-Feng and {Wang}, Yuan-Zhu and {Yuan}, Liang and {Wang}, Zhen-Han-Tao and {Hu}, Rui-Chong and {Wu}, Dong-Hong and {Yi}, Shuang-Xi and {Tang}, Qing-Wen and {Wei}, Jun-Jie and {Wu}, Xue-Feng and {Liang}, En-Wei},
        title = "{Stable case BB/BC mass transfer to form GW190425-like massive binary neutron star mergers}",
      journal = {\aap},
         year = 2024,
        month = nov,
       volume = {691},
          eid = {A214},
        pages = {A214},
          doi = {10.1051/0004-6361/202451444},
archivePrefix = {arXiv},
       eprint = {2409.10869},
 primaryClass = {astro-ph.SR},
       adsurl = {https://ui.adsabs.harvard.edu/abs/2024A&A...691A.214Q}
}

@ARTICLE{Qin_gw190425,
       author = {{Qin}, Ying and {Zhu}, Jin-Ping and {Meynet}, Georges and {Zhang}, Bing and {Wang}, Fa-Yin and {Shu}, Xin-Wen and {Song}, Han-Feng and {Wang}, Yuan-Zhu and {Yuan}, Liang and {Wang}, Zhen-Han-Tao and {Hu}, Rui-Chong and {Wu}, Dong-Hong and {Yi}, Shuang-Xi and {Tang}, Qing-Wen and {Wei}, Jun-Jie and {Wu}, Xue-Feng and {Liang}, En-Wei},
        title = "{Stable case BB/BC mass transfer to form GW190425-like massive binary neutron star mergers}",
      journal = {\aap},
         year = 2024,
        month = nov,
       volume = {691},
          eid = {A214},
        pages = {A214},
          doi = {10.1051/0004-6361/202451444},
archivePrefix = {arXiv},
       eprint = {2409.10869},
 primaryClass = {astro-ph.SR},
       adsurl = {https://ui.adsabs.harvard.edu/abs/2024A&A...691A.214Q}
}

@ARTICLE{Hachinger2012,
       author = {{Hachinger}, S. and {Mazzali}, P.~A. and {Taubenberger}, S. and {Hillebrandt}, W. and {Nomoto}, K. and {Sauer}, D.~N.},
        title = "{How much H and He is 'hidden' in SNe Ib/c? - I. Low-mass objects}",
      journal = {\mnras},
         year = 2012,
        month = may,
       volume = {422},
       number = {1},
        pages = {70-88},
          doi = {10.1111/j.1365-2966.2012.20464.x},
archivePrefix = {arXiv},
       eprint = {1201.1506},
 primaryClass = {astro-ph.SR},
       adsurl = {https://ui.adsabs.harvard.edu/abs/2012MNRAS.422...70H}
}

@ARTICLE{MLT1958,
       author = {{B{\"o}hm-Vitense}, E.},
        title = "{{\"U}ber die Wasserstoffkonvektionszone in Sternen verschiedener Effektivtemperaturen und Leuchtkr{\"a}fte. Mit 5 Textabbildungen}",
      journal = {\zap},
         year = 1958,
        month = jan,
       volume = {46},
        pages = {108},
       adsurl = {https://ui.adsabs.harvard.edu/abs/1958ZA.....46..108B}
}

@ARTICLE{Langer1983,
       author = {{Langer}, N. and {Fricke}, K.~J. and {Sugimoto}, D.},
        title = "{Semiconvective diffusion and energy transport}",
      journal = {\aap},
         year = 1983,
        month = sep,
       volume = {126},
       number = {1},
        pages = {207},
       adsurl = {https://ui.adsabs.harvard.edu/abs/1983A&A...126..207L}
}

@ARTICLE{Heger2000,
       author = {{Heger}, A. and {Langer}, N.},
        title = "{Presupernova Evolution of Rotating Massive Stars. II. Evolution of the Surface Properties}",
      journal = {\apj},
         year = 2000,
        month = dec,
       volume = {544},
       number = {2},
        pages = {1016-1035},
          doi = {10.1086/317239},
archivePrefix = {arXiv},
       eprint = {astro-ph/0005110},
 primaryClass = {astro-ph},
       adsurl = {https://ui.adsabs.harvard.edu/abs/2000ApJ...544.1016H}
}

@ARTICLE{Chaboyer1992,
       author = {{Chaboyer}, B. and {Zahn}, J. -P.},
        title = "{Effect of horizontal turbulent diffusion on transport by meridional circulation.}",
      journal = {\aap},
         year = 1992,
        month = jan,
       volume = {253},
        pages = {173-177},
       adsurl = {https://ui.adsabs.harvard.edu/abs/1992A&A...253..173C}
}

@ARTICLE{Heger1998,
       author = {{Heger}, A. and {Langer}, N.},
        title = "{The spin-up of contracting red supergiants}",
      journal = {\aap},
         year = 1998,
        month = jun,
       volume = {334},
        pages = {210-220},
archivePrefix = {arXiv},
       eprint = {astro-ph/9803005},
 primaryClass = {astro-ph},
       adsurl = {https://ui.adsabs.harvard.edu/abs/1998A&A...334..210H}
}

@ARTICLE{Langer1998,
       author = {{Langer}, N.},
        title = "{Coupled mass and angular momentum loss of massive main sequence stars}",
      journal = {\aap},
         year = 1998,
        month = jan,
       volume = {329},
        pages = {551-558},
       adsurl = {https://ui.adsabs.harvard.edu/abs/1998A&A...329..551L}
}

@ARTICLE{Maeder2000,
       author = {{Maeder}, A. and {Meynet}, G.},
        title = "{Stellar evolution with rotation. VI. The Eddington and Omega -limits, the rotational mass loss for OB and LBV stars}",
      journal = {\aap},
         year = 2000,
        month = sep,
       volume = {361},
        pages = {159-166},
archivePrefix = {arXiv},
       eprint = {astro-ph/0006405},
 primaryClass = {astro-ph},
       adsurl = {https://ui.adsabs.harvard.edu/abs/2000A&A...361..159M}
}

@ARTICLE{Zahn1977,
       author = {{Zahn}, J. -P.},
        title = "{Tidal friction in close binary systems.}",
      journal = {\aap},
         year = 1977,
        month = may,
       volume = {57},
        pages = {383-394},
       adsurl = {https://ui.adsabs.harvard.edu/abs/1977A&A....57..383Z}
}

@ARTICLE{Hut1981,
       author = {{Hut}, P.},
        title = "{Tidal evolution in close binary systems.}",
      journal = {\aap},
         year = 1981,
        month = jun,
       volume = {99},
        pages = {126-140},
       adsurl = {https://ui.adsabs.harvard.edu/abs/1981A&A....99..126H}
}

@ARTICLE{Hurley2002,
       author = {{Hurley}, Jarrod R. and {Tout}, Christopher A. and {Pols}, Onno R.},
        title = "{Evolution of binary stars and the effect of tides on binary populations}",
      journal = {\mnras},
         year = 2002,
        month = feb,
       volume = {329},
       number = {4},
        pages = {897-928},
          doi = {10.1046/j.1365-8711.2002.05038.x},
archivePrefix = {arXiv},
       eprint = {astro-ph/0201220},
 primaryClass = {astro-ph},
       adsurl = {https://ui.adsabs.harvard.edu/abs/2002MNRAS.329..897H}
}

@ARTICLE{Qin2018,
       author = {{Qin}, Y. and {Fragos}, T. and {Meynet}, G. and {Andrews}, J. and {S{\o}rensen}, M. and {Song}, H.~F.},
        title = "{The spin of the second-born black hole in coalescing binary black holes}",
      journal = {\aap},
         year = 2018,
        month = aug,
       volume = {616},
          eid = {A28},
        pages = {A28},
          doi = {10.1051/0004-6361/201832839},
archivePrefix = {arXiv},
       eprint = {1802.05738},
 primaryClass = {astro-ph.SR},
       adsurl = {https://ui.adsabs.harvard.edu/abs/2018A&A...616A..28Q}
}

@ARTICLE{Sciarini2024,
       author = {{Sciarini}, Luca and {Ekstr{\"o}m}, Sylvia and {Eggenberger}, Patrick and {Meynet}, Georges and {Fragos}, Tassos and {Song}, Han Feng},
        title = "{Dynamical tides in binaries: Inconsistencies in the implementation of Zahn's prescription}",
      journal = {\aap},
         year = 2024,
        month = jan,
       volume = {681},
          eid = {L1},
        pages = {L1},
          doi = {10.1051/0004-6361/202348424},
archivePrefix = {arXiv},
       eprint = {2312.08437},
 primaryClass = {astro-ph.SR},
       adsurl = {https://ui.adsabs.harvard.edu/abs/2024A&A...681L...1S}
}

@ARTICLE{Vink2001,
       author = {{Vink}, Jorick S. and {de Koter}, A. and {Lamers}, H.~J.~G.~L.~M.},
        title = "{Mass-loss predictions for O and B stars as a function of metallicity}",
      journal = {\aap},
         year = 2001,
        month = apr,
       volume = {369},
        pages = {574-588},
          doi = {10.1051/0004-6361:20010127},
archivePrefix = {arXiv},
       eprint = {astro-ph/0101509},
 primaryClass = {astro-ph},
       adsurl = {https://ui.adsabs.harvard.edu/abs/2001A&A...369..574V}
}

@ARTICLE{deJager,
       author = {{de Jager}, C. and {Nieuwenhuijzen}, H. and {van der Hucht}, K.~A.},
        title = "{Mass loss rates in the Hertzsprung-Russell diagram.}",
      journal = {\aaps},
         year = 1988,
        month = feb,
       volume = {72},
        pages = {259-289},
       adsurl = {https://ui.adsabs.harvard.edu/abs/1988A&AS...72..259D}
}

@ARTICLE{Nugis2000,
       author = {{Nugis}, T. and {Lamers}, H.~J.~G.~L.~M.},
        title = "{Mass-loss rates of Wolf-Rayet stars as a function of stellar parameters}",
      journal = {\aap},
         year = 2000,
        month = aug,
       volume = {360},
        pages = {227-244},
       adsurl = {https://ui.adsabs.harvard.edu/abs/2000A&A...360..227N}
}

@ARTICLE{Asplund2009,
       author = {{Asplund}, Martin and {Grevesse}, Nicolas and {Sauval}, A. Jacques and {Scott}, Pat},
        title = "{The Chemical Composition of the Sun}",
      journal = {\araa},
         year = 2009,
        month = sep,
       volume = {47},
       number = {1},
        pages = {481-522},
          doi = {10.1146/annurev.astro.46.060407.145222},
archivePrefix = {arXiv},
       eprint = {0909.0948},
 primaryClass = {astro-ph.SR},
       adsurl = {https://ui.adsabs.harvard.edu/abs/2009ARA&A..47..481A}
}

@ARTICLE{Vink2017,
       author = {{Vink}, Jorick S.},
        title = "{Winds from stripped low-mass helium stars and Wolf-Rayet stars}",
      journal = {\aap},
         year = 2017,
        month = nov,
       volume = {607},
          eid = {L8},
        pages = {L8},
          doi = {10.1051/0004-6361/201731902},
archivePrefix = {arXiv},
       eprint = {1710.02010},
 primaryClass = {astro-ph.SR},
       adsurl = {https://ui.adsabs.harvard.edu/abs/2017A&A...607L...8V}
}

@ARTICLE{Zhang2023,
       author = {{Zhang}, W.~T. and {Wang}, Z.~H.~T. and {Zhu}, J.-P. and {Hu}, R.-C. and {Shu}, X.~W. and {Tang}, Q.~W. and {Yi}, S.~X. and {Lyu}, F. and {Liang}, E.~W. and {Qin}, Y.},
        title = "{Super-Eddington accretion as a possible scenario to form GW190425}",
      journal = {\mnras},
         year = 2023,
        month = nov,
       volume = {526},
       number = {1},
        pages = {854-861},
          doi = {10.1093/mnras/stad2812},
archivePrefix = {arXiv},
       eprint = {2309.05189},
 primaryClass = {astro-ph.HE},
       adsurl = {https://ui.adsabs.harvard.edu/abs/2023MNRAS.526..854Z}
}

@ARTICLE{Wang2024,
       author = {{Wang}, Zhen-Han-Tao and {Hu}, Rui-Chong and {Qin}, Ying and {Zhu}, Jin-Ping and {Zhang}, Bing and {Yi}, Shuang-Xi and {Tang}, Qin-Wen and {Shu}, Xin-Wen and {Lyu}, Fen and {Liang}, En-Wei},
        title = "{A Channel to Form Fast-spinning Black Hole─Neutron Star Binary Mergers as Multimessenger Sources. II. Accretion-induced Spin-up}",
      journal = {\apj},
         year = 2024,
        month = apr,
       volume = {965},
       number = {2},
          eid = {177},
        pages = {177},
          doi = {10.3847/1538-4357/ad2fc1},
archivePrefix = {arXiv},
       eprint = {2401.17558},
 primaryClass = {astro-ph.HE},
       adsurl = {https://ui.adsabs.harvard.edu/abs/2024ApJ...965..177W}
}

@ARTICLE{Guo2024,
       author = {{Guo}, Yun-Lang and {Wang}, Bo and {Chen}, Wen-Cong and {Li}, Xiang-Dong and {Ge}, Hong-Wei and {Jiang}, Long and {Han}, Zhan-Wen},
        title = "{Electron-capture supernovae in NS + He star systems and the double neutron star systems}",
      journal = {\mnras},
         year = 2024,
        month = jun,
       volume = {530},
       number = {4},
        pages = {4461-4473},
          doi = {10.1093/mnras/stae1112},
archivePrefix = {arXiv},
       eprint = {2401.05103},
 primaryClass = {astro-ph.HE},
       adsurl = {https://ui.adsabs.harvard.edu/abs/2024MNRAS.530.4461G}
}

@ARTICLE{Wang2026,
       author = {{Wang}, Bo and {Liu}, Dongdong and {Guo}, Yunlang and {Han}, Zhanwen},
        title = "{The Formation of Electron-capture Supernovae: A Review}",
      journal = {Research in Astronomy and Astrophysics},
         year = 2026,
        month = mar,
       volume = {26},
       number = {3},
          eid = {032001},
        pages = {032001},
          doi = {10.1088/1674-4527/ae2d0e},
archivePrefix = {arXiv},
       eprint = {2509.25915},
 primaryClass = {astro-ph.HE},
       adsurl = {https://ui.adsabs.harvard.edu/abs/2026RAA....26c2001W}
}

@ARTICLE{Tauris2015,
       author = {{Tauris}, Thomas M. and {Langer}, Norbert and {Podsiadlowski}, Philipp},
        title = "{Ultra-stripped supernovae: progenitors and fate}",
      journal = {\mnras},
         year = 2015,
        month = aug,
       volume = {451},
       number = {2},
        pages = {2123-2144},
          doi = {10.1093/mnras/stv990},
archivePrefix = {arXiv},
       eprint = {1505.00270},
 primaryClass = {astro-ph.SR},
       adsurl = {https://ui.adsabs.harvard.edu/abs/2015MNRAS.451.2123T}
}

@ARTICLE{Choi2016,
       author = {{Choi}, Jieun and {Dotter}, Aaron and {Conroy}, Charlie and {Cantiello}, Matteo and {Paxton}, Bill and {Johnson}, Benjamin D.},
        title = "{Mesa Isochrones and Stellar Tracks (MIST). I. Solar-scaled Models}",
      journal = {\apj},
         year = 2016,
        month = jun,
       volume = {823},
       number = {2},
          eid = {102},
        pages = {102},
          doi = {10.3847/0004-637X/823/2/102},
archivePrefix = {arXiv},
       eprint = {1604.08592},
 primaryClass = {astro-ph.SR},
       adsurl = {https://ui.adsabs.harvard.edu/abs/2016ApJ...823..102C}
}

@ARTICLE{Tauris2012,
       author = {{Tauris}, T.~M. and {Langer}, N. and {Kramer}, M.},
        title = "{Formation of millisecond pulsars with CO white dwarf companions - II. Accretion, spin-up, true ages and comparison to MSPs with He white dwarf companions}",
      journal = {\mnras},
         year = 2012,
        month = sep,
       volume = {425},
       number = {3},
        pages = {1601-1627},
          doi = {10.1111/j.1365-2966.2012.21446.x},
archivePrefix = {arXiv},
       eprint = {1206.1862},
 primaryClass = {astro-ph.SR},
       adsurl = {https://ui.adsabs.harvard.edu/abs/2012MNRAS.425.1601T}
}

@ARTICLE{Landry2020,
       author = {{Landry}, Philippe and {Essick}, Reed and {Chatziioannou}, Katerina},
        title = "{Nonparametric constraints on neutron star matter with existing and upcoming gravitational wave and pulsar observations}",
      journal = {\prd},
         year = 2020,
        month = jun,
       volume = {101},
       number = {12},
          eid = {123007},
        pages = {123007},
          doi = {10.1103/PhysRevD.101.123007},
archivePrefix = {arXiv},
       eprint = {2003.04880},
 primaryClass = {astro-ph.HE},
       adsurl = {https://ui.adsabs.harvard.edu/abs/2020PhRvD.101l3007L}
}

@ARTICLE{Fryer2012,
       author = {{Fryer}, Chris L. and {Belczynski}, Krzysztof and {Wiktorowicz}, Grzegorz and {Dominik}, Michal and {Kalogera}, Vicky and {Holz}, Daniel E.},
        title = "{Compact Remnant Mass Function: Dependence on the Explosion Mechanism and Metallicity}",
      journal = {\apj},
         year = 2012,
        month = apr,
       volume = {749},
       number = {1},
          eid = {91},
        pages = {91},
          doi = {10.1088/0004-637X/749/1/91},
archivePrefix = {arXiv},
       eprint = {1110.1726},
 primaryClass = {astro-ph.SR},
       adsurl = {https://ui.adsabs.harvard.edu/abs/2012ApJ...749...91F}
}

@article{Song2023,
doi = {10.3847/1538-4357/acd6ee},
url = {https://doi.org/10.3847/1538-4357/acd6ee},
year = {2023},
month = {jul},
publisher = {The American Astronomical Society},
volume = {952},
number = {2},
pages = {156},
author = {Song, Cui-Ying and Liu, Tong},
title = {Long-duration Gamma-Ray Burst Progenitors and Magnetar Formation},
journal = {The Astrophysical Journal}
}

@ARTICLE{Spruit2002,
       author = {{Spruit}, H.~C.},
        title = "{Dynamo action by differential rotation in a stably stratified stellar interior}",
      journal = {\aap},
         year = 2002,
        month = jan,
       volume = {381},
        pages = {923-932},
          doi = {10.1051/0004-6361:20011465},
archivePrefix = {arXiv},
       eprint = {astro-ph/0108207},
 primaryClass = {astro-ph},
       adsurl = {https://ui.adsabs.harvard.edu/abs/2002A&A...381..923S}
}

@ARTICLE{Hu2022,
       author = {{Hu}, Rui-Chong and {Zhu}, Jin-Ping and {Qin}, Ying and {Zhang}, Bing and {Liang}, En-Wei and {Shao}, Yong},
        title = "{A Channel to Form Fast-spinning Black Hole-Neutron Star Binary Mergers as Multimessenger Sources}",
      journal = {\apj},
         year = 2022,
        month = apr,
       volume = {928},
       number = {2},
          eid = {163},
        pages = {163},
          doi = {10.3847/1538-4357/ac573f},
archivePrefix = {arXiv},
       eprint = {2201.09549},
 primaryClass = {astro-ph.HE},
       adsurl = {https://ui.adsabs.harvard.edu/abs/2022ApJ...928..163H}
}

@ARTICLE{Higgins2021,
       author = {{Higgins}, E.~R. and {Sander}, A.~A.~C. and {Vink}, J.~S. and {Hirschi}, R.},
        title = "{Evolution of Wolf-Rayet stars as black hole progenitors}",
      journal = {\mnras},
         year = 2021,
        month = aug,
       volume = {505},
       number = {4},
        pages = {4874-4889},
          doi = {10.1093/mnras/stab1548},
archivePrefix = {arXiv},
       eprint = {2105.12139},
 primaryClass = {astro-ph.SR},
       adsurl = {https://ui.adsabs.harvard.edu/abs/2021MNRAS.505.4874H}
}

@ARTICLE{Heger2005,
       author = {{Heger}, A. and {Woosley}, S.~E. and {Spruit}, H.~C.},
        title = "{Presupernova Evolution of Differentially Rotating Massive Stars Including Magnetic Fields}",
      journal = {\apj},
         year = 2005,
        month = jun,
       volume = {626},
       number = {1},
        pages = {350-363},
          doi = {10.1086/429868},
archivePrefix = {arXiv},
       eprint = {astro-ph/0409422},
 primaryClass = {astro-ph},
       adsurl = {https://ui.adsabs.harvard.edu/abs/2005ApJ...626..350H}
}

@ARTICLE{Petrovic2005,
       author = {{Petrovic}, J. and {Langer}, N. and {Yoon}, S.-C. and {Heger}, A.},
        title = "{Which massive stars are gamma-ray burst progenitors?}",
      journal = {\aap},
         year = 2005,
        month = may,
       volume = {435},
       number = {1},
        pages = {247-259},
          doi = {10.1051/0004-6361:20042545},
archivePrefix = {arXiv},
       eprint = {astro-ph/0504175},
 primaryClass = {astro-ph},
       adsurl = {https://ui.adsabs.harvard.edu/abs/2005A&A...435..247P}
}

@ARTICLE{Podsiadlowski2004,
       author = {{Podsiadlowski}, Ph. and {Langer}, N. and {Poelarends}, A.~J.~T. and {Rappaport}, S. and {Heger}, A. and {Pfahl}, E.},
        title = "{The Effects of Binary Evolution on the Dynamics of Core Collapse and Neutron Star Kicks}",
      journal = {\apj},
         year = 2004,
        month = sep,
       volume = {612},
       number = {2},
        pages = {1044-1051},
          doi = {10.1086/421713},
archivePrefix = {arXiv},
       eprint = {astro-ph/0309588},
 primaryClass = {astro-ph},
       adsurl = {https://ui.adsabs.harvard.edu/abs/2004ApJ...612.1044P}
}

@ARTICLE{Nomoto1984,
       author = {{Nomoto}, K.},
        title = "{Evolution of 8-10 solar mass stars toward electron capture supernovae. I - Formation of electron-degenerate O + NE + MG cores.}",
      journal = {\apj},
         year = 1984,
        month = feb,
       volume = {277},
        pages = {791-805},
          doi = {10.1086/161749},
       adsurl = {https://ui.adsabs.harvard.edu/abs/1984ApJ...277..791N}
}

@ARTICLE{Nomoto1987,
       author = {{Nomoto}, Ken'ichi},
        title = "{Evolution of 8--10 M$_{sun}$ Stars toward Electron Capture Supernovae. II. Collapse of an O + NE + MG Core}",
      journal = {\apj},
         year = 1987,
        month = nov,
       volume = {322},
        pages = {206},
          doi = {10.1086/165716},
       adsurl = {https://ui.adsabs.harvard.edu/abs/1987ApJ...322..206N}
}

@ARTICLE{Miyaji1980,
       author = {{Miyaji}, Shigeki and {Nomoto}, Ken'ichi and {Yokoi}, K{\~o}ichi and {Sugimoto}, Daiichiro},
        title = "{Supernova Triggered by Electron Captures}",
      journal = {\pasj},
         year = 1980,
        month = aug,
       volume = {32},
       number = {2},
        pages = {303-329},
          doi = {10.1093/pasj/32.2.303},
       adsurl = {https://ui.adsabs.harvard.edu/abs/1980PASJ...32..303M}
}

@ARTICLE{Dewi2002,
       author = {{Dewi}, J.~D.~M. and {Pols}, O.~R. and {Savonije}, G.~J. and {van den Heuvel}, E.~P.~J.},
        title = "{The evolution of naked helium stars with a neutron star companion in close binary systems}",
      journal = {\mnras},
         year = 2002,
        month = apr,
       volume = {331},
       number = {4},
        pages = {1027-1040},
          doi = {10.1046/j.1365-8711.2002.05257.x},
archivePrefix = {arXiv},
       eprint = {astro-ph/0201239},
 primaryClass = {astro-ph},
       adsurl = {https://ui.adsabs.harvard.edu/abs/2002MNRAS.331.1027D}
}

@ARTICLE{Dewi2003,
       author = {{Dewi}, J.~D.~M. and {Pols}, O.~R.},
        title = "{The late stages of evolution of helium star-neutron star binaries and the formation of double neutron star systems}",
      journal = {\mnras},
         year = 2003,
        month = sep,
       volume = {344},
       number = {2},
        pages = {629-643},
          doi = {10.1046/j.1365-8711.2003.06844.x},
archivePrefix = {arXiv},
       eprint = {astro-ph/0306066},
 primaryClass = {astro-ph},
       adsurl = {https://ui.adsabs.harvard.edu/abs/2003MNRAS.344..629D}
}

@ARTICLE{Ivanova2003,
       author = {{Ivanova}, N. and {Belczynski}, K. and {Kalogera}, V. and {Rasio}, F.~A. and {Taam}, R.~E.},
        title = "{The Role of Helium Stars in the Formation of Double Neutron Stars}",
      journal = {\apj},
         year = 2003,
        month = jul,
       volume = {592},
       number = {1},
        pages = {475-485},
          doi = {10.1086/375578},
archivePrefix = {arXiv},
       eprint = {astro-ph/0210267},
 primaryClass = {astro-ph},
       adsurl = {https://ui.adsabs.harvard.edu/abs/2003ApJ...592..475I}
}

@ARTICLE{Lattimer2001,
       author = {{Lattimer}, J.~M. and {Prakash}, M.},
        title = "{Neutron Star Structure and the Equation of State}",
      journal = {\apj},
         year = 2001,
        month = mar,
       volume = {550},
       number = {1},
        pages = {426-442},
          doi = {10.1086/319702},
archivePrefix = {arXiv},
       eprint = {astro-ph/0002232},
 primaryClass = {astro-ph},
       adsurl = {https://ui.adsabs.harvard.edu/abs/2001ApJ...550..426L}
}

@ARTICLE{Manchester2005,
       author = {{Manchester}, R.~N. and {Hobbs}, G.~B. and {Teoh}, A. and {Hobbs}, M.},
        title = "{The Australia Telescope National Facility Pulsar Catalogue}",
      journal = {\aj},
         year = 2005,
        month = apr,
       volume = {129},
       number = {4},
        pages = {1993-2006},
          doi = {10.1086/428488},
archivePrefix = {arXiv},
       eprint = {astro-ph/0412641},
 primaryClass = {astro-ph},
       adsurl = {https://ui.adsabs.harvard.edu/abs/2005AJ....129.1993M}
}

@ARTICLE{Peters1964,
       author = {{Peters}, P.~C.},
        title = "{Gravitational Radiation and the Motion of Two Point Masses}",
      journal = {Physical Review},
         year = 1964,
        month = nov,
       volume = {136},
       number = {4B},
        pages = {1224-1232},
          doi = {10.1103/PhysRev.136.B1224},
       adsurl = {https://ui.adsabs.harvard.edu/abs/1964PhRv..136.1224P}
}

@ARTICLE{Fuller2022,
       author = {{Fuller}, Jim and {Lu}, Wenbin},
        title = "{The spins of compact objects born from helium stars in binary systems}",
      journal = {\mnras},
         year = 2022,
        month = apr,
       volume = {511},
       number = {3},
        pages = {3951-3964},
          doi = {10.1093/mnras/stac317},
archivePrefix = {arXiv},
       eprint = {2201.08407},
 primaryClass = {astro-ph.HE},
       adsurl = {https://ui.adsabs.harvard.edu/abs/2022MNRAS.511.3951F}
}

@ARTICLE{Wu2022,
       author = {{Wu}, Samantha C. and {Fuller}, Jim},
        title = "{Extreme Mass Loss in Low-mass Type Ib/c Supernova Progenitors}",
      journal = {\apjl},
         year = 2022,
        month = nov,
       volume = {940},
       number = {1},
          eid = {L27},
        pages = {L27},
          doi = {10.3847/2041-8213/ac9b3d},
archivePrefix = {arXiv},
       eprint = {2210.10187},
 primaryClass = {astro-ph.HE},
       adsurl = {https://ui.adsabs.harvard.edu/abs/2022ApJ...940L..27W}
}

@ARTICLE{Hunter2007,
       author = {{Hunter}, John D.},
        title = "{Matplotlib: A 2D Graphics Environment}",
      journal = {Computing in Science and Engineering},
         year = 2007,
        month = jan,
       volume = {9},
       number = {3},
        pages = {90-95},
          doi = {10.1109/MCSE.2007.55},
       adsurl = {https://ui.adsabs.harvard.edu/abs/2007CSE.....9...90H}
}
\end{document}